\documentclass[twocolumn,aps,prb]{revtex4-1}

\usepackage{amsmath}  
\usepackage{amsfonts} 
\usepackage{graphicx} 

\begin{document}  

\title{Visualizing \emph{Interstellar}'s Wormhole}

\author{Oliver James}
\author{Eug\'enie von Tunzelmann}
\author{Paul Franklin}
\affiliation{Double Negative Ltd, 160 Great Portland Street, London W1W 5QA, UK}

\author{Kip S. Thorne}

\affiliation{California Institute of Technology, Pasadena, California 91125, USA}

\date{2 February 2015 -- American Journal of Physics, in press}

\begin{abstract}
Christopher Nolan's science fiction movie \emph{Interstellar} offers a variety of 
opportunities for students in elementary courses on general relativity theory.  
This paper describes such opportunities, including:  (i) At the motivational level,
the manner in which elementary relativity concepts underlie the wormhole 
visualizations seen in the movie.  (ii) At the briefest computational level, instructive
calculations with  simple but intriguing wormhole metrics, including, e.g., constructing
 embedding diagrams for the three-parameter wormhole that was used by
 our visual effects team and Christopher Nolan in scoping out possible
 wormhole geometries for the movie.  
(iii) Combining the proper reference frame of a camera with solutions
of the geodesic equation, to construct a light-ray-tracing map backward in time from a camera's local sky
to a wormhole's two celestial spheres.  (iv) Implementing this map, for example
in Mathematica, Maple or Matlab, and using that implementation to construct
images of what a camera sees when near or inside a wormhole.  (v) With the student's
implementation, exploring how the wormhole's three parameters influence what
the camera sees---which is precisely how Christopher Nolan, using our 
implementation, chose the parameters for \emph{Interstellar}'s wormhole.
(vi) Using the student's implementation,  exploring the wormhole's Einstein ring,
and particularly the peculiar motions of star images near the ring; and exploring
what it looks like to travel through a wormhole.

\end{abstract}

\maketitle

\section{Introduction}
\label{sec:Intro}

\subsection{The Context and Purposes of this paper}

In 1988, in connection with Carl Sagan's novel \textit{Contact}, \cite{Contact} later made into
a movie, \cite{ContactMovie} one of the authors published an article in this journal  about wormholes as a tool
for teaching general relativity (Morris and Thorne\cite{MorrisThorne}). 

This article is a follow-up, a quarter century later, in the context of Christopher
Nolan's movie \textit{Interstellar}\cite{Interstellar} and Kip Thorne's 
associated book
\emph{The Science of Interstellar}\cite{TSI}. Like \textit{Contact}, 
\textit{Interstellar} has real science built into its fabric, thanks
to a strong science commitment by the director, screenwriters, producers, and visual effects team, and thanks to Thorne's role as an executive producer. 

Although wormholes were central to the theme of \textit{Contact}  and to many
movies and TV shows since then, such as \textit{Star Trek} and \textit{Stargate}, none of
these have depicted correctly a wormhole as it would be seen by a nearby human.
\textit{Interstellar} is the first to do so.  The authors of this paper, together with Christopher Nolan who made key decisions, were responsible for that depiction.

This paper has two purposes: (i) To explain how \textit{Interstellar}'s wormhole images
were constructed and explain the decisions made on the way to their final form, and
(ii) to present this explanation in a way that may be useful to students and teachers in elementary courses on general relativity.

\subsection{The status of wormholes in the real universe}

Before embarking on these explanations, we  briefly describe physicists'
current understanding of wormholes, based on much research done  since 1988.  For a thorough and readable, but non-technical review, see the recent book
\textit{Time Travel and Warp Drives} by Allen Everett and Thomas Roman.\cite{EverettRoman}  For reviews that are more technical, see papers by Friedman and 
Higuchi\cite{FriedmanHiguchi} and by Lobo\cite{Lobo}.

In brief, physicists' current understanding is this:
\begin{itemize}
\item
There is no known mechanism for making wormholes, either naturally in our
universe or artificially by a highly advanced civilization, but there are speculations;
for example that wormholes in hypothetical quantum foam on the Planck scale, $\sqrt{G \hbar/c^3} \sim 10^{-35}$ m, might somehow be enlarged
to macroscopic size.\cite{MTU,EverettRoman}
\item
Any creation of a wormhole where initially there is none would require a change
in the topology of space, which would entail, in classical, non-quantum physics, both negative energy and closed timelike curves (the possibility of backward
time travel)---according to theorems by Frank Tipler and Robert Geroch.\cite{FriedmanHiguchi}   It is likely the laws of physics forbid this. Likely
but not certain.
\item
A wormhole will pinch off so quickly that nothing can travel through it, unless it has
``exotic matter'' at its throat---matter (or fields) that, at least in some reference frames, has
negative energy density.  Although such negative energy density is permitted by the laws of physics
(e.g.\ in the Casimir effect, the electromagnetic field between two highly conducting
plates), there are quantum inequalities that limit the amount of negative energy that can
be collected in a small region of space and how long it can be there; and these 
\emph{appear} to place severe limits on the sizes of traversable wormholes (wormholes
through which things can travel at the speed of light or slower).\cite{EverettRoman} The
implications of these inequalities are not yet fully clear, but it seems likely that,
after some strengthening,
they will prevent macroscopic wormholes  like the one in \textit{Interstellar} from
staying open long enough for a spaceship to travel through.  Likely, but not certain.

\item
The research leading to these conclusions has been performed ignoring the possibility
that our universe, with its four spacetime dimensions, resides in a higher
dimensional \textit{bulk} with one or more large extra dimensions, the kind 
of bulk envisioned in \textit{Interstellar}'s ``fifth dimension.''  Only a little is known about how such a bulk might influence the
existence of traversable wormholes, but one intriguing thing is clear: Properties of the
bulk can, at least in principle, hold a wormhole open without any need for 
exotic matter in our four dimensional universe (our ``brane'').\cite{Lobo}  But the
words ``in principle'' just hide our great ignorance about our universe in higher
dimensions.
\end{itemize}

In view of this current understanding, it seems very unlikely to us that traversable wormholes
exist naturally in our universe, and the prospects for highly advanced civilizations
to make them artificially are also pretty dim. 

Nevertheless, the distances from our solar system to others are so huge that there is 
little hope, with  rocket technology,  for humans to travel to other stars in the next
century or two;\cite{OtherStars} so wormholes, quite naturally, have become a staple
of science fiction.

And, as Thorne envisioned in 1988,\cite{MorrisThorne} wormholes have also become a pedagogical
tool in elementary courses on general relativity---e.g., in the textbook by James
Hartle.\cite{Hartle}

\subsection{The genesis of our research on wormholes}

This paper is a collaboration between Caltech physicist Kip Thorne, and 
computer graphics artists
at \textit{Double Negative Visual Effects} in London.  
We came together in May 2013, when Christopher Nolan asked us to collaborate on building, for \emph{Interstellar}, realistic images of a wormhole, and also a fast spinning black hole and its accretion disk, with ultra-high (IMAX) resolution and smoothness.  We saw this  not only as an opportunity to bring realistic wormholes and black holes
into the Hollywood arena, but also  an opportunity to create images of wormholes and black holes for relativity and astrophysics research.

Elsewhere\cite{HoleLens} we describe the simulation code that we wrote for this: DNGR for
``Double Negative Gravitational Renderer'', and
the black-hole and accretion-disk images we generated with it, and also some new insights into gravitational lensing by 
black holes that it has revealed.  In this paper we focus on wormholes---which are
much easier to model mathematically than \textit{Interstellar}'s fast spinning black hole, and are
far more easily incorporated into elementary courses on general relativity.

In our modelling of \emph{Interstellar}'s wormhole, we pretended we were engineers in some arbitrarily
advanced civilization, and  that  the laws of physics place no constraints
on the wormhole  geometries our construction crews can build.  (This is almost certainly false;
the quantum inequalities mentioned above, or other physical laws, likely place strong constraints on wormhole geometries, if wormholes are allowed at all---but we know so little about those constraints that we chose to ignore them.)  In this spirit, we wrote down the spacetime 
metrics for candidate wormholes for the movie, and then proceeded to visualize them.

\subsection{Overview of this paper}

We begin in Sec.\ \ref{sec:WormholeMetrics} by presenting the spacetime metrics for several wormholes
and visualizing them with embedding diagrams --- most importantly, the three-parameter
``Dneg wormhole'' metric used in our work on the movie \emph{Interstellar}.  Then we
discuss adding a Newtonian-type gravitational potential to our Dneg metric, to produce
the gravitational pull that Christopher Nolan wanted, and the potential's unimportance
for making wormhole images.

In Sec\ \ref{sec:Mapping} we describe how light rays, traveling backward in time from a camera to the wormhole's two celestial spheres, generate a map that can be used
to produce images of the wormhole and of objects seen through or around it; and we discuss
our implementations of that map to make the images seen in \emph{Interstellar}.  In the 
Appendix we present a fairly simple computational procedure by which students can
generate their own map and thence their own images.

In Sec.\ \ref{sec:ParameterInfluences} we use our own implementation of the map to describe the influence of
the Dneg wormhole's three parameters on what the camera sees. 
 
Then in Secs.\ \ref{sec:InterstellarWormhole} and \ref{sec:ThroughWormhole}, we discuss
Christopher Nolan's use of these kinds of implementations to choose the
parameter values for \emph{Interstellar}'s wormhole;  we discuss the resulting
wormhole images that appear in \emph{Interstellar}, including that wormhole's
Einstein ring, which can be explored by watching the movie or its trailers, or in
students' own implementations of the ray-tracing map; and we discuss images
made by a camera travelling through the wormhole, that do not appear in the movie.

Finally in Sec.\ \ref{sec:Conclusion} we present brief conclusions.

Scattered throughout the paper are suggestions of calculations and projects
for students in elementary courses on general relativity.  And 
throughout, as is common in relativity, we use ``geometrized units'' in which Newton's gravitational
constant $G$ and the speed of light $c$ are set equal to unity, so time is measured
in length units, 1 s = $c \times$1 s = $2.998 \times 10^8$ m; and mass is expressed 
in length units:  1 kg = $(G/c^2)\times $1 kg $= 0.742 \times 10^{-27}$
 m; and the mass
of the Sun is 1.476 km.

\section{Spacetime Metrics for Wormholes, and Embedding Diagrams}
\label{sec:WormholeMetrics}

In general relativity, the curvature of spacetime can be expressed, mathematically,
in terms of a spacetime metric.  In this section we review a simple example of
this: the metric for an \emph{Ellis wormhole}; and then we discuss the metric for
the 
\emph{Double Negative (Dneg) wormhole}  that we designed for  \emph{Interstellar}.

\subsection{The Ellis wormhole}
\label{subsec:EllisWormhole}

In 1973 Homer Ellis\cite{Ellis} introduced the following metric for a hypothetical wormhole,
which he called a ``drainhole'':\cite{Ellis-MT}
\begin{equation}
ds^2 = -dt^2 + d\ell^2 + r^2 (d\theta^2 +\sin^2\theta\, d\phi^2)\;, 
\label{eq:GeneralMetric}
\end{equation}
where $r$ is a function of the coordinate $\ell$ given by
\begin{equation}
r(\ell)=\sqrt{\rho^2 + \ell^2}\;,
\label{eq:rEllis}
\end{equation}
and $\rho$ is a constant.

As always in general relativity, one does not need to be told anything about the coordinate system in order to figure out the spacetime geometry described by the metric; the metric by itself tells us everything.  Deducing everything is
a good exercise for students.  Here is how we do so:

First, in $-dt^2$ the minus sign tells us that $t$, at fixed $\ell$, $\theta$, $\phi$, 
increases in a timelike direction; and the absence of any factor multiplying $-dt^2$ tells
us that $t$ is, in fact, proper time (physical time) measured by somebody at rest in
the spatial, $\{\ell,\theta,\phi\}$ coordinate system.

Second, the expression $ r^2(d\theta^2 +\sin^2\theta\, d\phi^2)$ is the familiar metric for the
surface of a sphere with circumference $2 \pi r$ and surface area $4\pi r^2$, written in spherical polar coordinates $\{\theta,\phi\}$, so the Ellis wormhole
must be spherically symmetric.  As we would in flat space, we
shall use the name ``radius'' for the sphere's circumference divided by $2\pi$, i.e.\
for $r$.  For the Ellis wormhole, this radius is $r=\sqrt{\rho^2 + \ell^2}$.

Third, from the plus sign in front of $d\ell^2$ we infer that $\ell$ is a spatial coordinate; and since there are no cross terms $d\ell d\theta$ or $d\ell d\phi$, the coordinate lines of
constant $\theta$ and $\phi$, with increasing $\ell$, must be radial lines; and since
 $d\ell^2$ has no multiplying coefficient, $\ell$ must be the proper distance (physical) distance traveled in that radial direction.
 
Fourth, when $\ell$ is large and negative, the radii of spheres $r=\sqrt{\rho^2+\ell^2}$ is large
and approximately equal to $|\ell|$.  When $\ell$ increases to zero, $r$ decreases to
its minimum value $\rho$.  And when $\ell$ increases onward to a very large value,
$r$ increases once again, becoming approximately $\ell$.  This tells us that the metric
represents a wormhole with throat radius $\rho$, connecting two asymptotically flat regions of space, $\ell\rightarrow - \infty$ and $\ell\rightarrow+\infty$.

\begin{figure}
\begin{center}
\includegraphics[width=0.95\columnwidth]{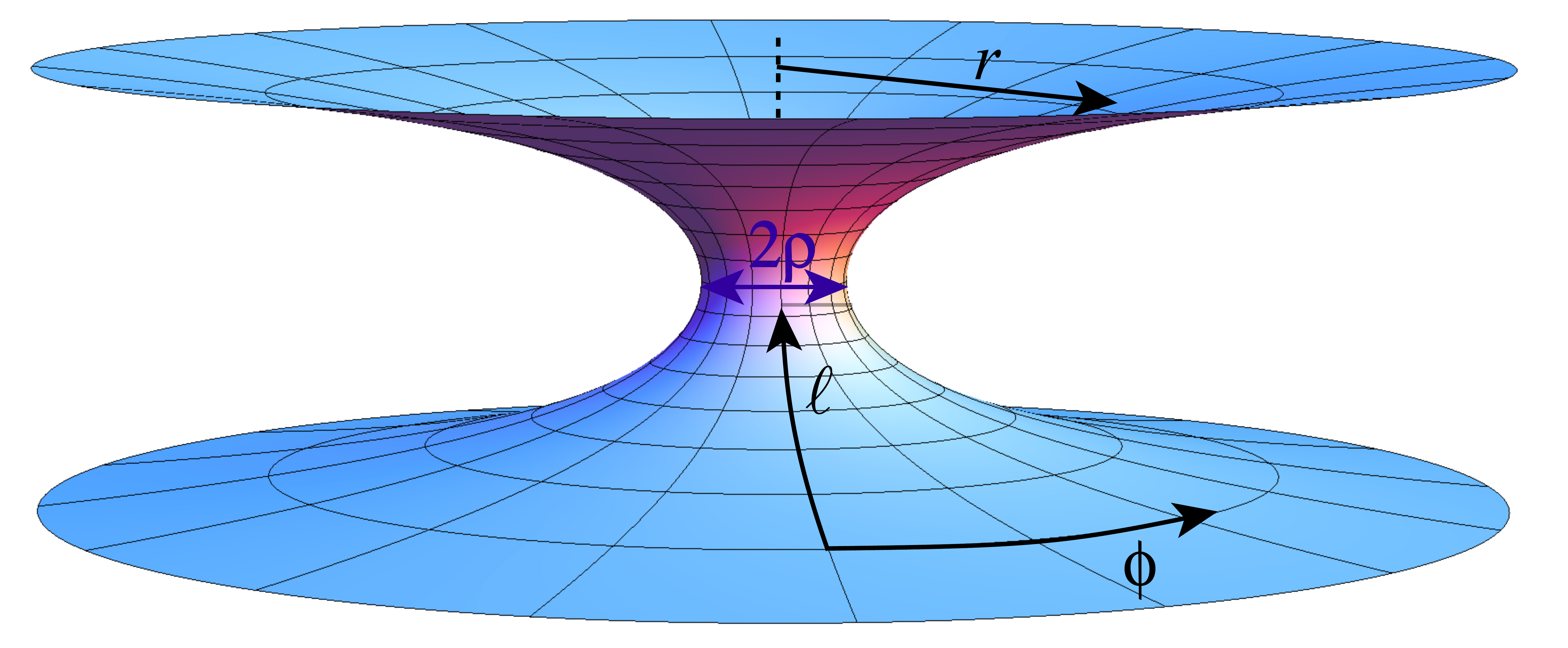}
\end{center}
\caption{Embedding diagram for the Ellis wormhole: the wormhole's  two-dimensional equatorial plane embedded in three of the bulk's four spatial dimensions.}
\label{fig-EllisWormhole}
\end{figure}

In Hartle's textbook,\cite{Hartle} a number of illustrative calculations are carried out using Ellis's
wormhole metric as an example.  The most interesting is a computation,
in Sec.\ 7.7,  of what
the two-dimensional equatorial surfaces (surfaces with constant $t$ and $\theta = 
\pi/2$) look like when embedded in a flat 3-dimensional space, the
\emph{embedding space}.
Hartle shows that equatorial surfaces have
the form shown in Fig.\ \ref{fig-EllisWormhole}---a form familiar from popular
accounts of wormholes.  

Figure \ref{fig-EllisWormhole} is called an ``embedding
diagram'' for the wormhole.  We discuss embedding diagrams further in 
Sec.\ \ref{subsec:DnegEmbedding} below, in the context of 
our Dneg wormhole.

Thomas M\"uller and colleagues \cite{Muller} have visualized an Ellis wormhole in various environments by methods similar to those that we lay out below.  

\subsection{The Double Negative three-parameter wormhole}
\label{subsec:DnegWormhole}

The Ellis wormhole was not an appropriate starting point for our {\it Interstellar} work.  Christopher Nolan, the movie's director, wanted to see how the wormhole's visual appearance depends on its shape, so the shape had to be adjustable, which was not the case for the Ellis wormhole.

So for \emph{Interstellar} we designed a wormhole
with three free shaping parameters and produced images of what a camera orbiting the wormhole
would see for various values of the parameters.  Christopher Nolan and Paul
Franklin, the leader of our Dneg effort,  then discussed the images; and based on them,
Nolan chose the parameter values for the movie's wormhole.

In this section we 
explain our three-parameter Double Negative (Dneg) wormhole in three steps: First, a variant with just two
parameters (the length and radius of the wormhole's interior) and with sharp 
transitions from its
interior to its exteriors; then a variant with a third parameter, called the \emph{lensing
length}, that smooths the transitions; and finally a variant in which we add a gravitational pull.

\subsubsection{Wormhole with sharp transitions}

\begin{figure}
\begin{center}
\includegraphics[width=0.95\columnwidth]{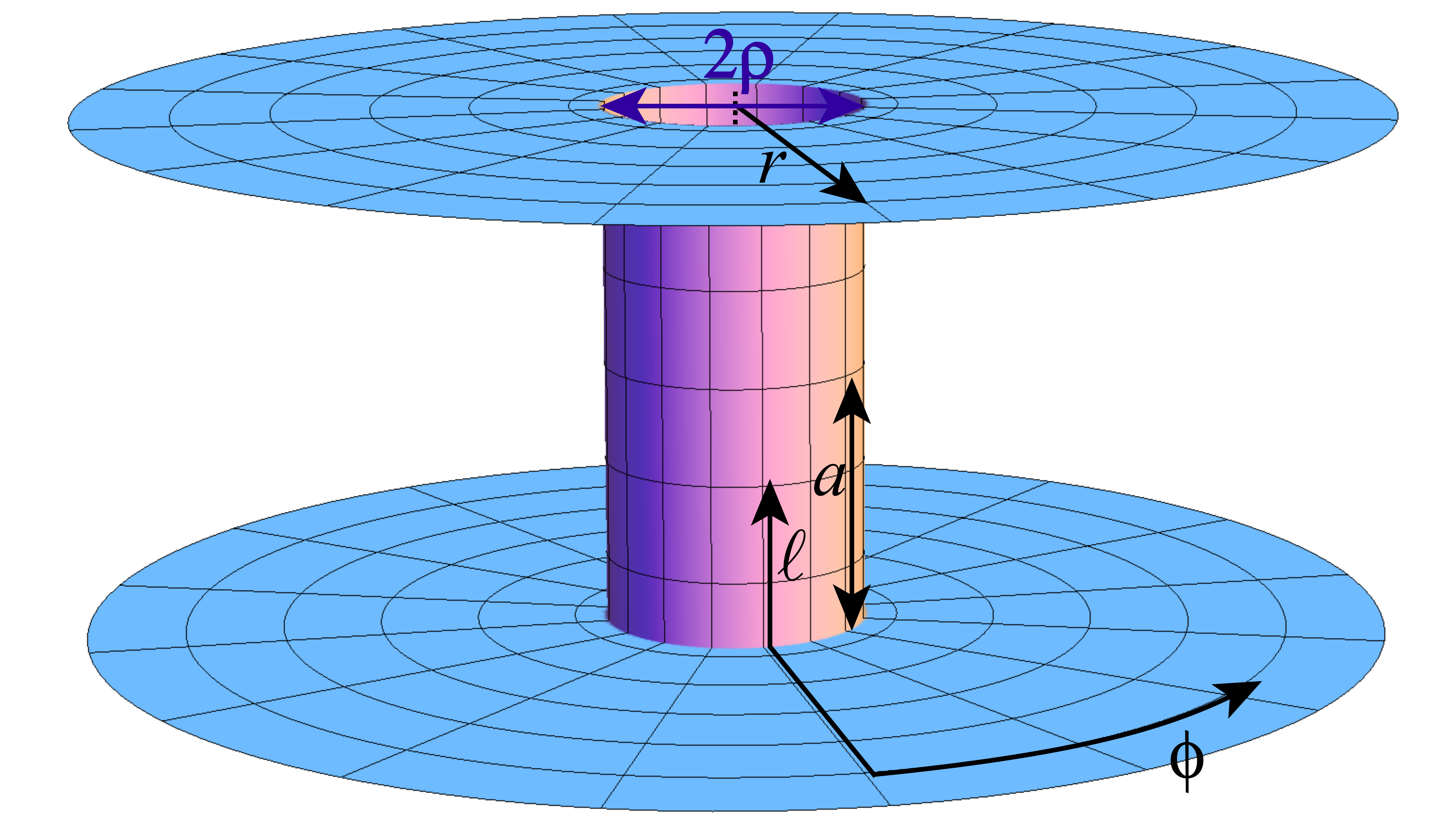}
\end{center}
\caption{Embedding diagram for the wormhole with sharp transition,
Eqs.\ (\ref{eq:GeneralMetric}) and (\ref{eq:rDnegS}).} 
\label{fig-STWormhole}
\end{figure}

Our wormhole with sharp transitions is a simple cylinder of length $2a$, whose cross sections are spheres, all with the same radius $\rho$; this cylinder is joined at its ends onto flat three-dimensional
spaces with balls of radius $\rho$ removed.  This wormhole's embedding diagram 
is Fig.\  \ref{fig-STWormhole}.  As always, the embedding diagram has one spatial
dimension removed, so the wormhole's cross sections appear as circles rather than
spheres.

Using the same kinds of spherical polar coordinates as for the Ellis wormhole above,
the spacetime metric has the general wormhole form (\ref{eq:GeneralMetric}) with
 \begin{eqnarray}
 r(\ell)&=&\rho  \quad \textrm{for the wormhole interior, } |\ell|\le a\;,
  \label{eq:STr} 
\label{eq:rDnegS}\\
 &=& |\ell|-a+\rho \quad \textrm{for the wormhole exterior, } |\ell|> a\;.
\nonumber 
 \end{eqnarray}
 
 \subsubsection{Dneg wormhole without gravity}
 
Our second step is to smooth the transitions between the wormhole interior
$|\ell|<a$ 
(the cylinder) and the two external universes $|\ell| > a$.  As we shall see, the smoothed
transitions give rise to gravitational lensing (distortions) of the star field behind each wormhole mouth.  Such gravitational lensing is a big deal in astrophysics and 
cosmology these days; see, e.g., the Gravitational Lensing Resource Letter\cite{GLRL}; and, as we discuss in Sec.\ \ref{subsec:EinsteinRing}, it shows up in a rather weird way,
in \emph{Interstellar}, near the edges of the wormhole image.  

Somewhat arbitrarily, we chose to make the transition have approximately the
same form as that from the throat (horizon) of a nonspinning black hole to the
external universe in which the hole lives.  Such a hole's metric (the ``Schwarzschild
metric'') has a form that is most simply written using radius $r$ as the outward coordinate
rather than proper distance $\ell$:  
\begin{equation}
ds^2 = -(1-2\mathcal M/r)dt^2 + {dr^2 \over 1-2\mathcal M/r} + r^2 (d\theta^2 + \sin^2\theta\, d\phi^2)\;,
\label{eq:SchMetric}
\end{equation}
where $\mathcal M$ is the black hole's mass.
Comparing the spatial part of this metric ($t = $constant) with our general wormhole
metric (\ref{eq:GeneralMetric}), we see that $d\ell = \pm dr/\sqrt{1-2\mathcal M/r}$, which
can easily be integrated to obtain the proper distance traveled as a function of
radius, $\ell(r)$.  What we want, however, is $r$ as a function of $\ell$, and we
want it in an analytic form that is easy to work with; so for our Dneg wormhole,
we choose a fairly simple analytic function that is roughly the same as the Schwarzschild
$r(\ell)$:  

Outside the wormhole's cylindrical interior, we chose
\begin{subequations}
\begin{eqnarray}
r &=& \rho+ {2\over\pi} \int_0^{|\ell|-a} \arctan\left({2\xi\over \pi \mathcal M}\right)d\xi
\label{eq:rDnegA} \\
&=& \rho + \mathcal M \left[ x \arctan x - \frac12\ln(1+x^2)\right]\;,  \;\;
\textrm{for } |\ell|>a\;, 
\label{eq:rellDneg}
\nonumber
\end{eqnarray}
where
\begin{equation}
x\equiv {2(|\ell|-a)\over \pi \mathcal M}\;.
\label{eq:rDnegB}
\end{equation}
(Students might want to compare this graphically with the inverse of the Schwarzschild
$\ell = \int dr/\sqrt{1-2 {\cal M}/r}$, plotting, e.g., $r-\rho$ for our wormhole as a function of $|\ell|-a$; and $r-2M$
of Schwarzschild as a function of distance from the Schwarzschild horizon $r=2M$.)
Within the wormhole's cylindrical interior, we chose, of course,
\begin{equation}
r= \rho \quad \textrm{for } |\ell|< a\;.
\label{eq:rDnegC}
\end{equation}
\label{eq:rDneg}
\end{subequations}
These equations (\ref{eq:rDneg}) for $r(\ell)$, together with our general wormhole
metric (\ref{eq:GeneralMetric}), describe the spacetime geometry of the Dneg
wormhole without gravity.

For the Schwarzschild metric, the throat radius $\rho$ is equal to twice the
black hole's mass (in geometrized units), $\rho = 2 \mathcal M$.  For our 
Dneg wormhole
we choose the two parameters $\rho$ and $\mathcal M$ to be independent: 
they represent the wormhole's radius and the gentleness of the transition from
the wormhole's cylindrical interior to its asymptotically flat exterior.

We shall refer to the ends of the cylindrical interior, $\ell = \pm a$, as 
the wormhole's \emph{mouths}.  They are spheres with circumferences $2\pi\rho$.

\subsubsection{Embedding diagrams for the Dneg wormhole}
\label{subsec:DnegEmbedding}

We construct embedding diagrams for the Dneg wormhole (and any other
spherical wormhole) by comparing the spatial metric of the wormhole's
two-dimensional equatorial surface $ds^2 = d\ell^2 + r^2(\ell) d\phi^2$ with
the spatial metric of the embedding space. Doing so is a good exercise
for students.  For the embedding space we choose cylindrical
coordinates with the symmetry axis along  the wormhole's center line.  Then (as
in Figs.\ \ref{fig-EllisWormhole} and \ref{fig-STWormhole}), the embedding
space and
the wormhole share the same radial coordinate $r$ and angular coordinate
$\phi$, so with $z$ the embedding-space height above the wormhole's midplane, 
the embedding-space metric is $ds^2 = dz^2 + dr^2 + r^2 d\phi^2$.  
Equating this to
the wormhole metric, we see that\cite{HartleEmbed} $dz^2 + dr^2 = d\ell^2$, which gives us 
an equation for the height $z$ of the wormhole surface as a function of 
distance $\ell$ through the wormhole:
\begin{equation}
z(\ell) = \int_0^\ell \sqrt{1-(dr/d\ell')^2} d\ell'\;.
\label{eq:EmbeddingEqn}
\end{equation}

By inserting the Dneg radius function (\ref{eq:rDneg}) into this expression
and performing the integral numerically, we obtain the wormhole shapes
shown in Fig.\ \ref{fig-Dnegwormhole1pp5} and Figs.\ \ref{fig:DnegVary_a} and
\ref{fig:DnegVary_W} below.

\begin{figure}
\begin{center}
\includegraphics[width=0.95\columnwidth]{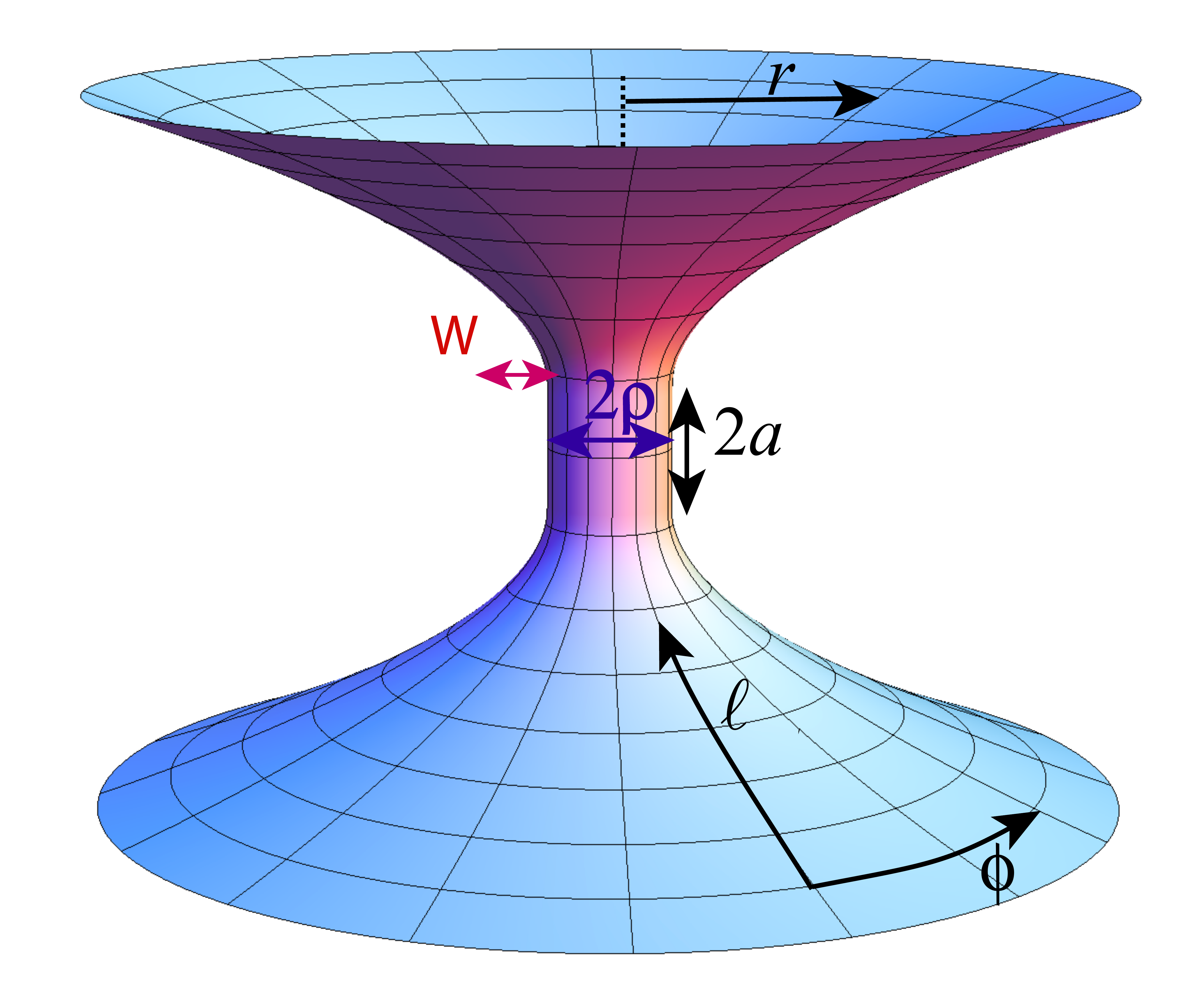}
\end{center}
\caption{Embedding diagram for the Dneg wormhole with parameters
$a/\rho =1$ (length $2a$ of cylindrical section equal to its diameter $2\rho$) and 
$\mathcal M/\rho = 0.5$, which corresponds to a lensing width
$\mathcal W/\rho = 0.715$.}
\label{fig-Dnegwormhole1pp5}
\end{figure}

The actual shape of this embedding diagram depends on two dimensionless ratios
of the Dneg metric's three parameters: the wormhole's length-to-diameter ratio
$2a/2\rho = a/\rho$, and its ratio
$\mathcal M/\rho$.  For chosen values of these ratios, the wormhole's size
is then fixed by its interior radius $\rho$, which Christopher Nolan chose
to be one kilometer in \emph{Interstellar}, so with the technology of the movie's 
era the wormhole's gravitational lensing of our galaxy's star field can be 
seen from Earth, but barely so.\cite{NolanRadius} 

In the embedding diagram of Fig.\ 
\ref{fig-Dnegwormhole1pp5}, instead of depicting
$\mathcal M$, we depict the lateral distance
$\mathcal W$ in the embedding space, over which the wormhole's surface changes
from vertical to 45 degrees.  This $\mathcal W$ is related to $\mathcal M$ by \cite{LensingLength}
\begin{equation}
\mathcal W = 1.42953...\, \mathcal M
\label{eq:LensingLength}
\end{equation}
We call this $\mathcal W$ the wormhole's \emph{Lensing width}, and we
often use it in place of $\mathcal M$ as the wormhole's third parameter.

\subsubsection{Dneg wormhole with gravity}

Christopher Nolan asked for the movie's spacecraft Endurance to travel along a trajectory that gives enough time for the audience to view the wormhole up close before Cooper, the pilot, initiates descent into the wormhole's mouth.  Our Double Negative team designed such a trajectory, which required that the wormhole have a gravitational acceleration of order the Earth's, $\sim 10$m/s$^2$, or less.  This is 
so weak that it can be 
described accurately by a Newtonian gravitational potential $\Phi$ of magnitude $|\Phi|\ll c^2 = 1$ (see below), that shows up in the
time part of the metric.  More specifically, we modify the wormhole's metric 
(\ref{eq:GeneralMetric}) to read
\begin{equation}
ds^2 = -(1+2\Phi)dt^2 + d\ell^2 + r^2 (d\theta^2 +\sin^2\theta\, d\phi^2)\;. 
\label{eq:MetricWithGravity}
\end{equation}
The sign of $\Phi$ is
negative (so the wormhole's gravity will be attractive), and spherical symmetry dictates that it be a function only of $\ell$.  

According to the equivalence principle, the gravitational acceleration experienced by
a particle at rest outside or inside the wormhole (at fixed spatial coordinates
$\{\ell,\theta,\phi\} = $ constant) is the negative of that particle's 4-acceleration.  Since the
4-acceleration is orthogonal to the particle's 4-velocity, which points in the time direction, its 
gravitational acceleration is purely spatial in the coordinate system $\{t,\ell,\theta,\phi\}$.  
It is a nice exercise for students to compute the particle's 4-acceleration and thence
its gravitational acceleration.  The result, aside from negligible fractional corrections of order $|\Phi|$, is 
\begin{equation}
\mathbf g = - (d\Phi/d\ell)\, \mathbf e_{\hat \ell}\;,
\label{eq:GravitationalAcceleration}
\end{equation}
where $\mathbf e_{\hat \ell}$ is the unit vector pointing in the radial direction.  
Students may have seen an equation analogous to (\ref{eq:MetricWithGravity})
when space is nearly flat, and a calculation in that case which yields 
Eq.\ (\ref{eq:GravitationalAcceleration}) for $\bf g$ (e.g.\ Sec.\ 
6.6 of Hartle\cite{Hartle}).  Although for the wormhole metric 
(\ref{eq:MetricWithGravity}), with $r$ given by Eqs.\ (\ref{eq:rDneg}) or 
(\ref{eq:rEllis}), space is far from flat, Eq.\ (\ref{eq:GravitationalAcceleration}) is still
true---a deep fact that students would do well to absorb and  generalize.

It is reasonable to choose the gravitational acceleration $g = |{\bf g}| = |d\Phi/d\ell |$ 
to fall off as $\sim 1/(\textrm{distance})^2$  as we move away from the wormhole mouth; or at least faster than $\sim 1/(\textrm{distance})$.  Integrating $g =  |d\Phi/d\ell |$ 
radially and using this rapid falloff, the student can deduce that the
magnitude of $\Phi$ is of order $g$ times the wormhole's radius $\rho$.  With a gravitational acceleration $g = |{\bf g}|\lesssim 10$  m/s$^2$ and $\rho = 1$ km, this gives 
$|\Phi| \sim |{\bf g}|\rho \lesssim 10^4$(m/s)$^2 \sim10^{-12}$.  
Here we have divided by the speed
of light squared to bring this into our geometrized units.  

Such a tiny gravitational potential corresponds to a slowing of time near the wormhole
by the same small amount, no more than a part in $10^{12}$ [cf.\ the time part of the metric
(\ref{eq:MetricWithGravity})].  This is so small as to be utterly unimportant in the movie,
and so small that, when computing the propagation of light rays through the wormhole,
to ultrahigh accuracy we can ignore $\Phi$ and use the Dneg metric without gravity. We shall do so.

\section{Mapping a Wormhole's Two Celestial Spheres onto a Camera's Sky}
\label{sec:Mapping}

\subsection{Foundations for the Map}
\label{subsec:MapFoundations}

\begin{figure}[b!]
\begin{center}
\includegraphics[width=0.95\columnwidth]{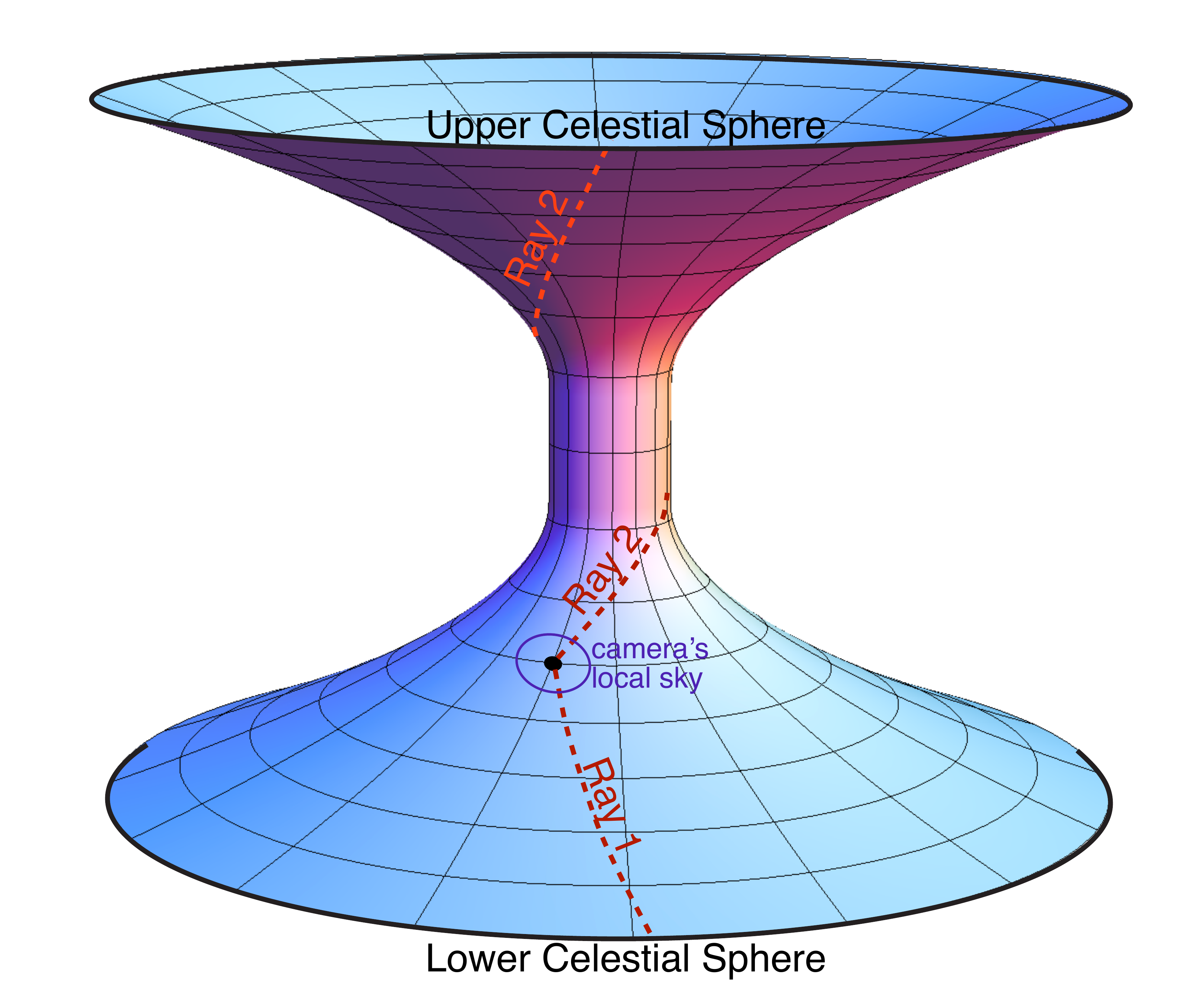}
\end{center}
\vskip-1.5pc
\caption{Embedding diagram showing light rays 1 and 2 that carry light from a wormhole's lower and upper celestial spheres, to a camera.  The celestial spheres are incorrectly depicted close to the wormhole; they actually are very far away, and we idealize them as at $\ell = \pm\infty$. 
}
\label{fig:WormholeMapping}
\end{figure}

A camera inside or near a wormhole receives light rays from light sources and uses them to create images.  In this paper we shall assume, for simplicity, that all the light sources are far from the wormhole, so far that we can idealize them as lying on ``celestial spheres" at $\ell \rightarrow - \infty$ (lower celestial sphere; Saturn side of the wormhole in the movie \emph{Interstellar}) and $\ell \rightarrow + \infty$  (upper celestial sphere; Gargantua side in \emph{Interstellar}); see Fig.\ \ref{fig:WormholeMapping}.  (Gargantua
is a supermassive black hole in the movie that humans visit.)  Some light rays carry light from the lower celestial sphere to the camera's local sky (e.g.\ Ray 1 in Fig.\ \ref{fig:WormholeMapping}); others carry light from the upper celestial sphere to the camera's local sky (e.g.\ Ray 2).  Each of these rays is a null geodesic through the wormhole's spacetime.

\begin{figure}[b!]
\begin{center}
\includegraphics[width=1.0\columnwidth]{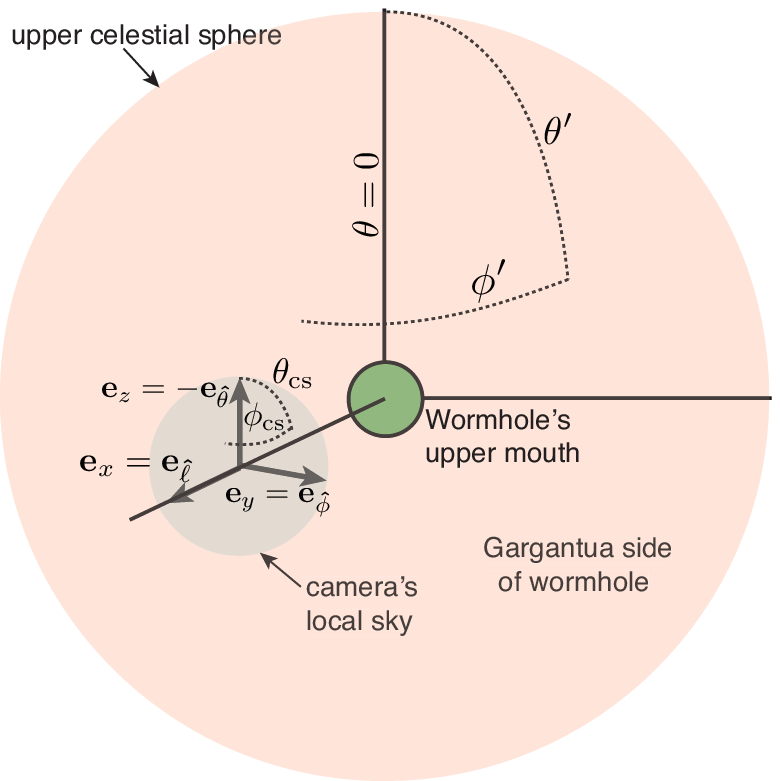}
\vskip2pc
\includegraphics[width=1.0\columnwidth]{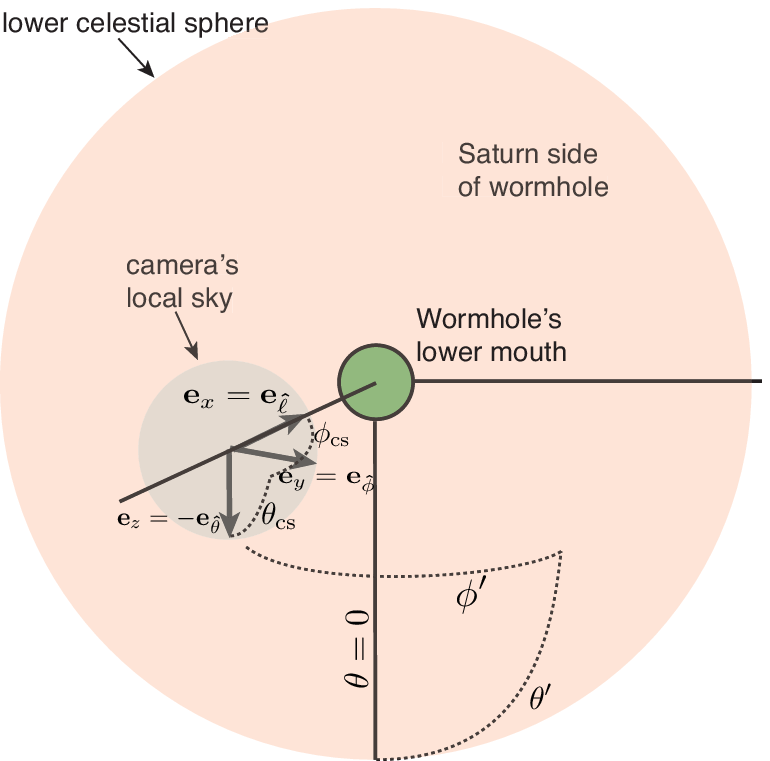}
\end{center}
\vskip-1.5pc
\caption{The two sides of the wormhole, with a camera on each side at $\theta_c=\pi/2$
(equatorial plane), $\phi_c=0$, and $\ell_c > a$ on the Gargantua side; $\ell_c<-a $ on the
Saturn side. 
}
\label{fig:WH-TwoSides}
\end{figure}

On each celestial sphere, we  set up spherical polar coordinates $\{\theta',\phi'\}$,
which are the limits of the spherical polar coordinates $\{\theta,\phi\}$ as 
$\ell \rightarrow \pm \infty$.  We draw these two celestial spheres in 
Fig.\  \ref{fig:WH-TwoSides}, a diagram of the three dimensional space around each wormhole mouth, with the curvature of space not shown.   Notice that we choose to draw the north polar
axes $\theta=0$ pointing away from each other and the south polar axes $\theta=\pi$
pointing toward each other.  This is rather arbitrary, but it feels comfortable to us when we
contemplate the  embedding diagram of Fig.\ \ref{fig:WormholeMapping}.

We assume the camera moves at speeds very low compared to light speed (as it does in \emph{Interstellar}), so relativistic
aberration and doppler shifts are unimportant,  Therefore, when computing images the camera
makes, we can treat the camera as at rest in the $\{\ell,\theta,\phi\}$ 
coordinate system.

We can think of the camera as having a local sky, on which there are spherical
polar coordinates $\{\theta_{\rm cs}, \phi_{\rm cs}\}$ (``cs'' for camera sky; not to
be confused with celestial sphere!); Fig.\ \ref{fig:WH-TwoSides}. In more technical language, $\{\theta_{\rm cs}, \phi_{\rm cs}\}$  are spherical polar coordinates for the tangent space at the 
camera's location.  

A light ray that heads backward in time from the camera (e.g.\ Ray 1 or 2 in Fig.\ 
\ref{fig:WormholeMapping}), traveling in the 
$\{\theta_{\rm cs}, \phi_{\rm cs}\}$ direction, ultimately winds up at location
$\{\theta',\phi'\}$ on one of the wormhole's two celestial spheres.  It brings to
$\{\theta_{\rm cs}, \phi_{\rm cs}\}$ on the camera's sky an image of whatever
was at $\{\theta',\phi'\}$ on the celestial sphere.  

This means that the key to making images of what the camera sees is a ray-induced map
from the camera's sky to the celestial spheres:  $\{\theta',\phi',s\}$ as a function of
$\{\theta_{\rm cs},\phi_{\rm cs}\}$, where the parameter $s$ tells us which celestial
sphere the backward light ray reaches: the upper one ($s=+$) or the lower one ($s=-$).  

In the Appendix we sketch a rather simple computational procedure by which students can 
compute this map and then, using it, can construct images of wormholes and their surroundings; and we describe a Mathematica implementation of this procedure by this paper's computationally challenged author Kip Thorne.

\subsection{Our DNGR Mapping and Image Making}
\label{subsec:DNGR}

To produce the IMAX images needed for \emph{Interstellar}, 
at Double Negative we developed a much more sophisticated implementation of the map within within a computer code that we call DNGR\cite{HoleLens} (Double Negative Gravitational Renderer).  In DNGR, we
use ray bundles (light beams) to do the mapping rather than just light rays.  We begin with a circular
light beam about one pixel in size at the camera and trace it backward in time to its
origin on a 
celestial sphere using the ray equations (\ref{eq:RE}), plus the general relativistic  equation of geodesic deviation, which evolves the beam's size and shape.  At the celestial sphere, the beam is an ellipse, often
 highly eccentric.  We integrate up the image data within that ellipse to 
deduce the light traveling into the camera's circular pixel.
We also do spatial filtering to smooth artifacts and time filtering to mimic the
behavior of a movie camera (when the image is changing rapidly), and we sometimes
add lens flare to mimic the effects of light scattering and diffraction in a movie camera's
lens.  

Elsewhere\cite{HoleLens} we give some details of these various ``bells and whistles'', for a camera orbiting a black hole rather than a wormhole.  They are essentially
the same for a wormhole. 

However, fairly nice images can be produced without any of these bells and whistles,
using the simple procedure described in the Appendix, 
and thus are within  easy reach of students in an elementary course on
general relativity.

\section{The Influence of the Wormhole's  parameters on what the camera sees}
\label{sec:ParameterInfluences}

For Christopher Nolan's perusal in choosing \emph{Interstellar}'s wormhole parameters, we used our map to make
images of the galaxy in which the black hole Gargantua resides, as viewed from the 
Saturn side of the wormhole; see below.  But for this paper, and the book\cite{TSI} that
Thorne has written about the science of \emph{Interstellar}, we find it more instructive,
pedagogically, to show images of Saturn and its rings as seen through the wormhole
from the Gargantua side.  This section is a more quantitative version of a discussion
of this in Chap.\ 15 of that book.\cite{TSI}

\begin{figure}[h!]
\begin{center}
\includegraphics[width=1.0\columnwidth]{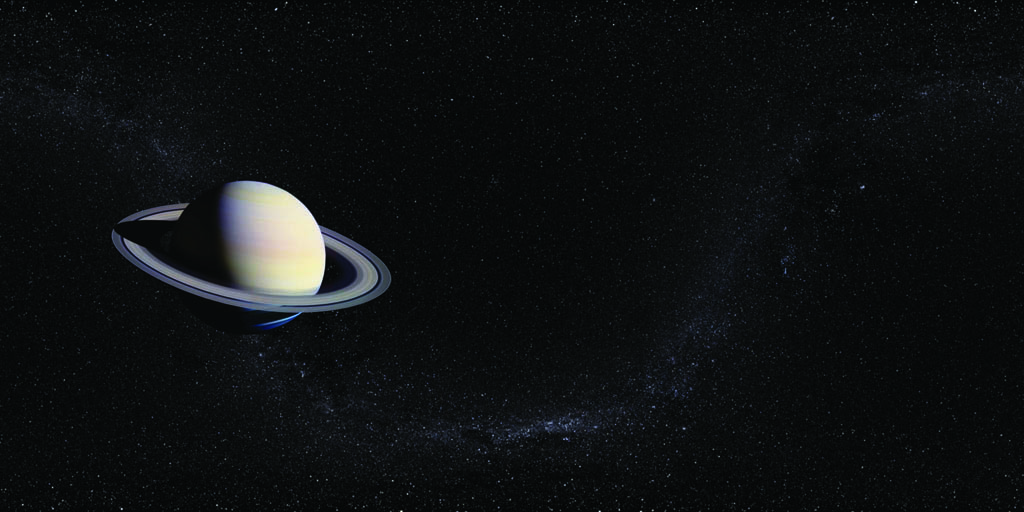}
\includegraphics[width=1.0\columnwidth]{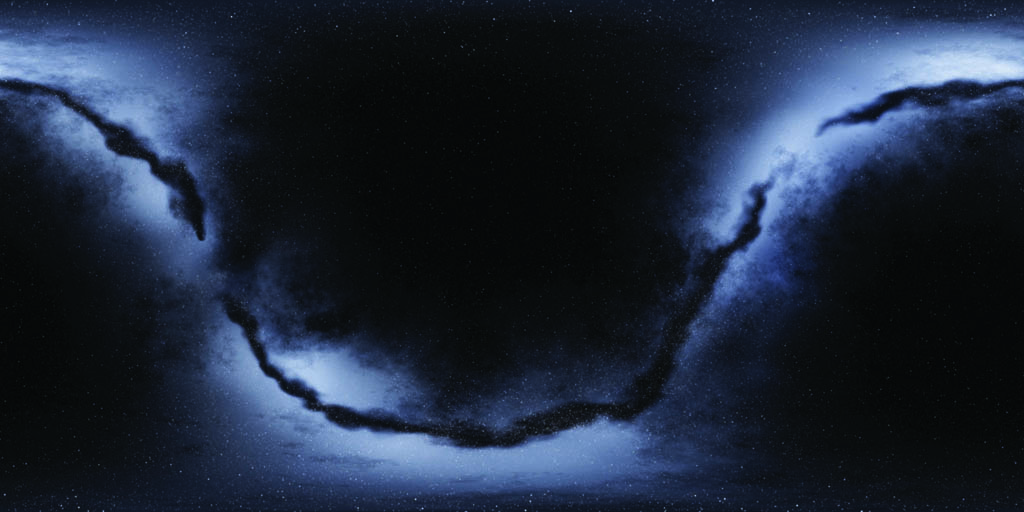}
\end{center}
\vskip-1.5pc
\caption{(a) The image of Saturn placed on the lower celestial sphere of Fig.\ 
\ref{fig:WH-TwoSides}. [From a composition of Cassini data by Mattias Malmer \cite{Malmer}.]  (b) The star-field image placed on the upper celestial sphere. [Created
by our Double Negative artistic team].  These images are available in high resolution,
for use by students, at \protect\url{http://www.dneg.com/dneg_vfx/wormhole}. }
\label{fig:CelestialSpheres}
\end{figure}

Figure \ref{fig:CelestialSpheres} shows the simple Saturn image that we placed on the 
lower celestial sphere of Fig.\  \ref{fig:WH-TwoSides}, and a star field that we placed on the upper
celestial sphere (the Gargantua side of the wormhole).   
Both images are mapped from the celestial sphere onto a flat rectangle with
azimuthal angle $\phi$ running horizontally and polar angle $\theta$ vertically.
In computer graphics, this type of image is known as a \emph{longitude-latitude map.}
\cite{BLINN}

\subsection{Influence of the Wormhole's Length}
\label{subsec:WHlength}

\begin{figure}
\begin{center}
\includegraphics[width=1.0\columnwidth]{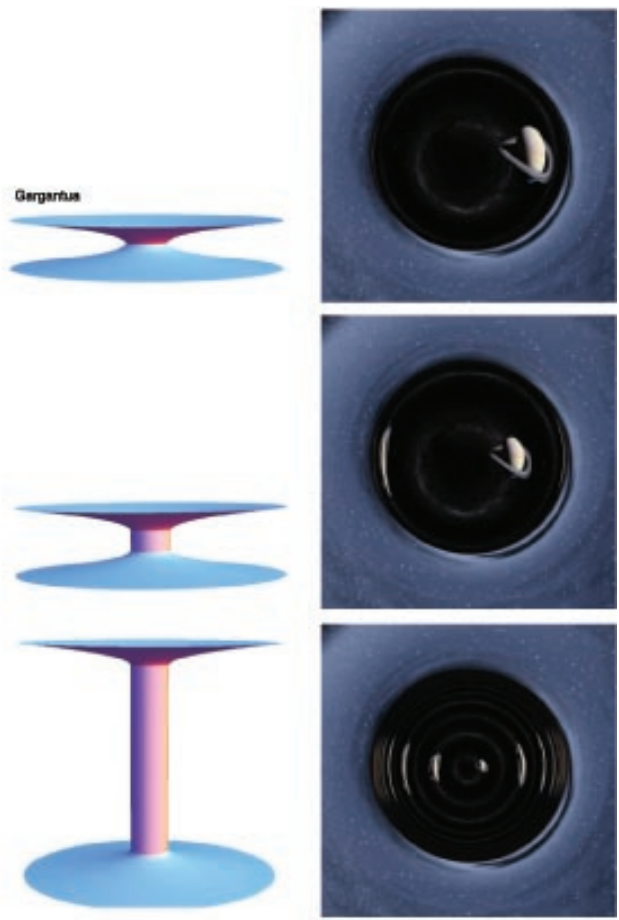}
\end{center}
\vskip-1.5pc
\caption{Images of Saturn on the camera sky, as seen through the wormhole,
for small lensing width, $\mathcal W=0.05 \rho$ and various wormhole 
lengths, from top to bottom, $2a/\rho = 0.01, 1, 10$.  The camera is 
at $\ell = 6.25\rho + a$; i.e., at a distance $6.25\rho$ from the wormhole's mouth---the edge of its cylindrical interior.
[Adapted from Fig.\ 15.2 of  \emph{The Science of Interstellar}\cite{TSI}, 
and used by
permission of W. W. Norton \& Company, Inc.
TM \& $\copyright$ 2015 Warner Bros. Entertainment Inc. (s15), and Kip 
Thorne.
\emph{Interstellar} and all related characters and elements are
trademarks of and \copyright Warner Bros. Entertainment Inc. (s15).
The images on the right 
may be used under the terms
of the Creative Commons Attribution-NonCommercial-NoDerivs 3.0 (CC BY-NC-ND
3.0) license. Any further distribution of these images must maintain attribution to the
author(s) and the title of the work, journal citation and DOI. You may not use the
images for commercial purposes and if you remix, transform or build upon the images,
you may not distribute the modified images.]
}
\label{fig:DnegVary_a}
\end{figure}

In Fig.\  \ref{fig:DnegVary_a} we explore the influence of the wormhole's 
length on the camera-sky image produced by these two celestial spheres.  
Specifically, we hold the wormhole's lensing width fixed at a fairly small value,
$\mathcal W = 0.05 \rho$, and we vary the wormhole's length
from $2a = 0.01\rho$ (top picture), to $2a=\rho$ (middle picture), to $2a=10\rho$
(bottom picture).  

Because Saturn and its rings are white and the sky around it is black, while the star field
on the Gargantua side of the wormhole is blue, we can easily identify the edge of
the wormhole
mouth as the transition from black-and-white to blue.  (The light's colors are preserved as 
the light travels near and through the wormhole because we have assumed the wormhole's
gravity is weak, $|\Phi|\ll 1$; there are no significant gravitational frequency shifts.)

\begin{figure}
\begin{center}
\includegraphics[width=1.0\columnwidth]{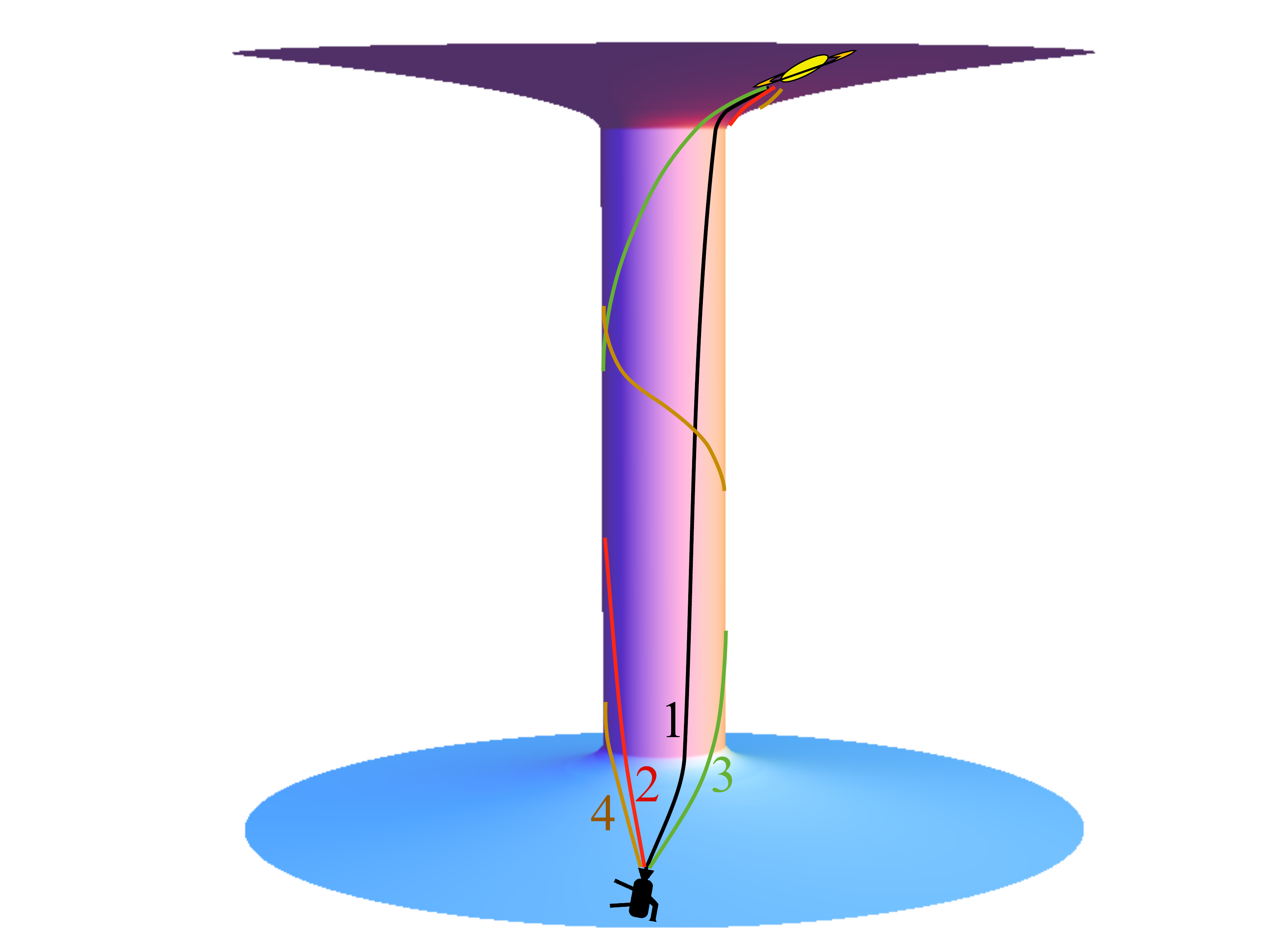}
\end{center}
\vskip-1.5pc
\caption{Light rays that travel from Saturn, though the Dneg wormhole, to the camera, 
producing the images in Fig.\ \ref{fig:DnegVary_a}. [Adapted from Fig.\ 15.3 of  \emph{The Science of Interstellar}\cite{TSI}.]}
\label{fig:Rays}
\end{figure}

Through a short wormhole (top), the camera sees a large distorted image of Saturn
nearly filling the right half of the wormhole mouth.  This is the primary image, carried by
light rays that travel on the shortest possible paths through the wormhole from
Saturn to camera, such as the black path in Fig.\ \ref{fig:Rays}.  There is also a very thin, lenticular, secondary image of
Saturn, barely discernable, near the  left edge of the wormhole mouth.  It is brought
to the camera by light rays that travel around the left side of the wormhole (e.g.
path 2 in Fig.\ \ref{fig:Rays})---a 
longer route than for the primary image.  The lenticular
structure at the lower right is blue, so it is a secondary gravitationally lensed image of the blue star field that resides on
the camera's side of the wormhole.

As the wormhole is lengthened (middle of Fig.\ \ref{fig:DnegVary_a}), the primary and 
secondary images move inward and shrink in size.  A lenticular tertiary image emerges 
from the mouth's right edge, carried by rays like 3 in Fig.\ \ref{fig:Rays} that wrap 
around the wormhole once; and a fourth faint, lenticular image emerges from the
left side, carried by rays like 4 that wrap around the wormhole in the opposite
direction, one and a half times.  

As the wormhole is lengthened more and more (bottom of  Fig.\ \ref{fig:DnegVary_a}),
the existing images shrink and move inward toward the mouth's center, and new
images emerge, one after another, from the right then left then right... sides of the
mouth.

For a short wormhole, all these images were already present, very near the wormhole's
edge; but they were so thin as to be unresolvable.  Lengthening the wormhole moved
them inward and made them thick enough to see.

\subsection{Influence of the Wormhole's Lensing Width}
\label{subsec:WHLensingWidth}

\begin{figure}
\begin{center}
\includegraphics[width=1.0\columnwidth]{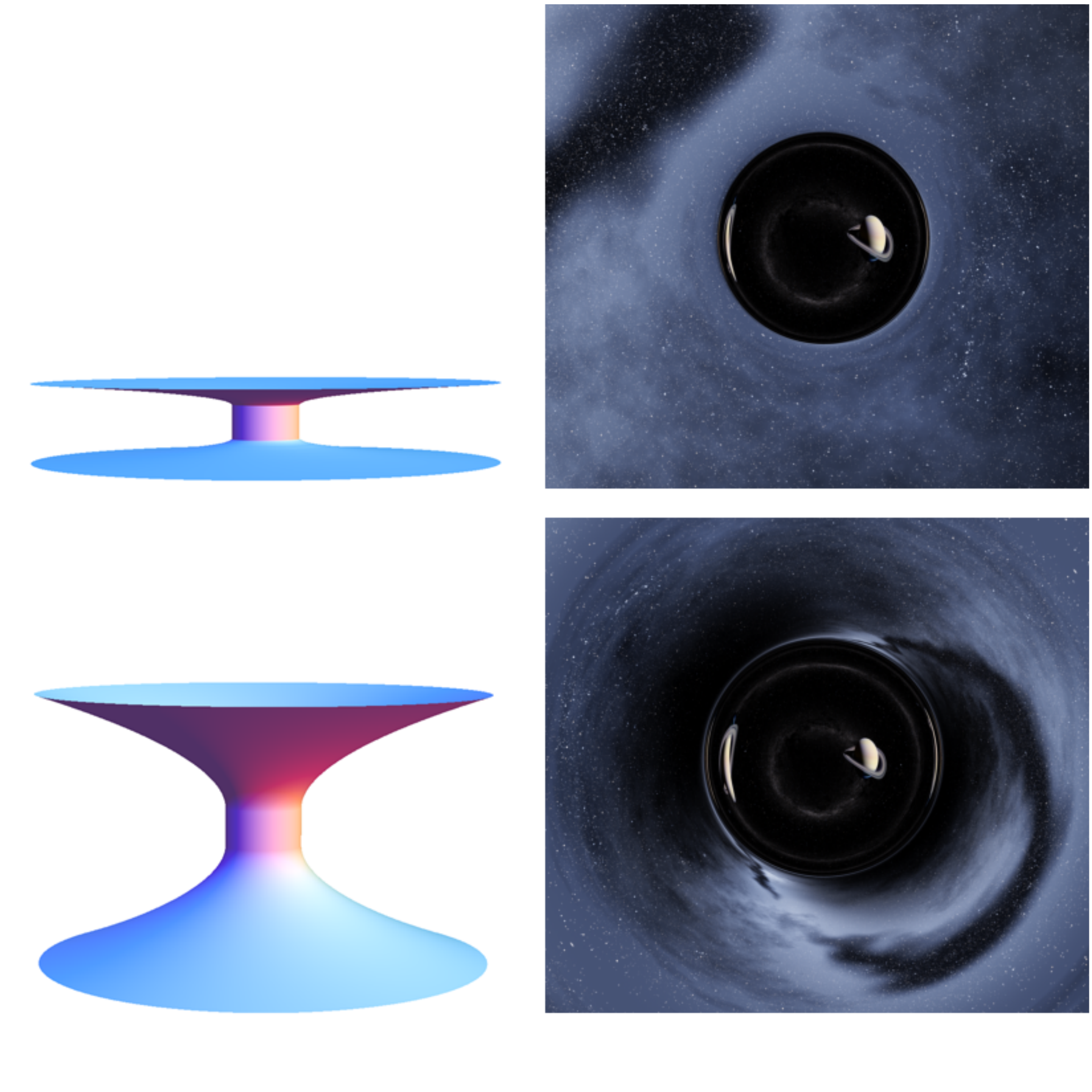}
\end{center}
\vskip-1.5pc
\caption{Images of Saturn on the camera sky, as seen through a  wormhole
with fixed length equal to the wormhole radius, $2a = \rho$, and for two
lensing widths: $\mathcal W=0.014 \rho$ (top) and $\mathcal W = 0.43$ (bottom). 
 [Adapted from Fig.\ 15.4 of  \emph{The Science of Interstellar}\cite{TSI}, 
and used by
permission of W. W. Norton \& Company, Inc.
 TM \& $\copyright$ Warner Bros. Entertainment Inc. (s15), and Kip 
Thorne.  The images on the right 
may be used under the terms
of the Creative Commons Attribution-NonCommercial-NoDerivs 3.0 (CC BY-NC-ND
3.0) license. Any further distribution of these images must maintain attribution to the
author(s) and the title of the work, journal citation and DOI. You may not use the
images for commercial purposes and if you remix, transform or build upon the images,
you may not distribute the modified images.]}
\label{fig:DnegVary_W}
\end{figure}

In Fig.\ \ref{fig:DnegVary_W} we explore the influence of the wormhole's lensing width on 
what the camera sees.  We hold its length fixed and fairly small: equal to its radius, $2a=\rho$.

For small lensing width $\mathcal W = 0.014\rho$ (top), the transition from the wormhole's cylindrical interior to its asymptotically flat exterior is quite sharp; so,
not surprisingly, the camera sees an exterior, blue star field that extends with little
distortion right up to the edge of the wormhole mouth.

By contrast, when the lensing width is larger, $\mathcal W = 0.43\rho$ (bottom), 
the external star field is greatly distorted by gravitational lensing.   The dark cloud
on the upper left side of the wormhole is enlarged and pushed out
of the cropped picture, and we see a big secondary image
of the cloud on the wormhole's lower right and a tertiary image on its upper left.  We also see lensing of the wormhole
mouth itself: it is enlarged; and lensing of the image that comes through the wormhole
from the Saturn side.  The lenticular secondary image of Saturn near the mouth's left edge is thickened, while the primary image is shrunken a bit and moved inward
to make room for a new tertiary image on the right.

Students could check their wormhole imaging code by trying to reproduce 
one or more images from Figs.\ \ref{fig:DnegVary_a} and \ref{fig:DnegVary_W},
using the images in Fig.\ \ref{fig:CelestialSpheres} on their celestial spheres. Having
done so, they could further explore the influence of the wormhole parameters on
the images the camera sees.

\section{\emph{Interstellar}'s Wormhole}
\label{sec:InterstellarWormhole} 

After reviewing images analogous to Figs.\ \ref{fig:DnegVary_a} and 
\ref{fig:DnegVary_W},  but with Saturn replaced by the stars and nebulae of
\emph{Interstellar}'s distant galaxy (the galaxy
on the Gargantua side of the wormhole; Fig.\  \ref{fig:InterstellarGalaxy}), Christopher Nolan
made his choice for the parameters of \emph{Interstellar}'s wormhole.

He chose a very short wormhole: length $2a = 0.01 \rho$ as in the top
panel of Fig.\ \ref{fig:DnegVary_a}; for greater lengths the multiple images
would be confusing to a mass audience.  And he chose a modest lensing 
width: $\mathcal W = 0.05 \rho$ also as in the top panel of Fig.\ \ref{fig:DnegVary_a}
and in between the two lensing widths of Fig.\ \ref{fig:DnegVary_W}.
This gives enough gravitational lensing to be interesting (see below), but far
less lensing than for a black hole, thereby enhancing the visual distinction between
\emph{Interstellar}'s wormhole and its black hole Gargantua.

\subsection{\emph{Interstellar}'s Distant Galaxy}
\label{subsec:InterstellarGalaxy}

For \emph{Interstellar}, a team under the leadership of authors Paul Franklin
and Eug\'enie von Tunzelmann constructed images of the distant galaxy through
a multistep process:

The distant end of the wormhole was imagined to be in the distant galaxy and
closer to its center than we are to the center of our Milky Way.
Consequently the view of the surrounding galaxy must be recognisably different
from the view we have from Earth: larger and brighter nebulae, more dense dust,
with brighter and more numerous visible stars. This view was created as an artistic task.

Nebulae were painted (by texture artist Zoe Lord), using a combination of space
photography and imagination, covering a range of colour palettes. These were
combined with layers of painted bright space dust and dark, silhouetted dust
channels, to create a view of the galaxy with as much visual depth and complexity as possible.

Star layout was achieved by taking real star data as seen from Earth and
performing various actions to make the view different: the brightest stars
were removed from the data set (to avoid recognisable constellations) and
the brightnesses of all the other stars were increased and shuffled. The
result was a believably natural-looking star layout which was unrecognisable
compared to our familiar view of the night sky from Earth.

\begin{figure}
\begin{center}
\includegraphics[width=1.0\columnwidth]{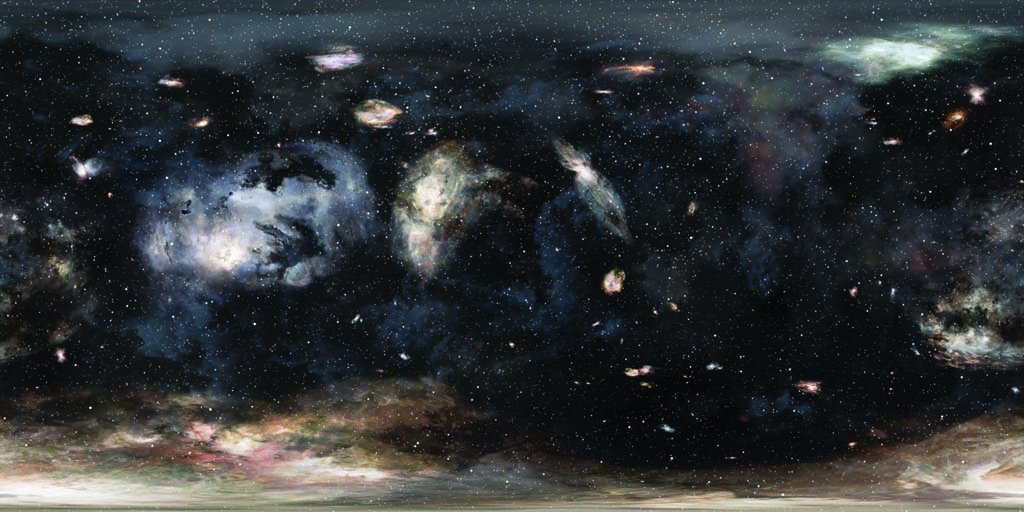}
\end{center}
\vskip-1.5pc
\caption{An image of stars and nebulae in  \emph{Interstellar}'s distant galaxy (the
galaxy on the Gargantua side of the wormhole), created by our Double Negative
artistic team. This image is available in high resolution,
for use by students, at \protect\url{http://www.dneg.com/dneg_vfx/wormhole}. }
\label{fig:InterstellarGalaxy}
\end{figure}

Figure \ref{fig:InterstellarGalaxy} is one of our distant-galaxy images, showing nebulae,
space dust and stars.

\subsection{View through \textit{\textbf{Interstellar}}'s Wormhole}
\label{subsec:InterstellarWormhole}

\begin{figure}[h!]
\vskip1.5pc
\begin{center}
\includegraphics[width=1.0\columnwidth]{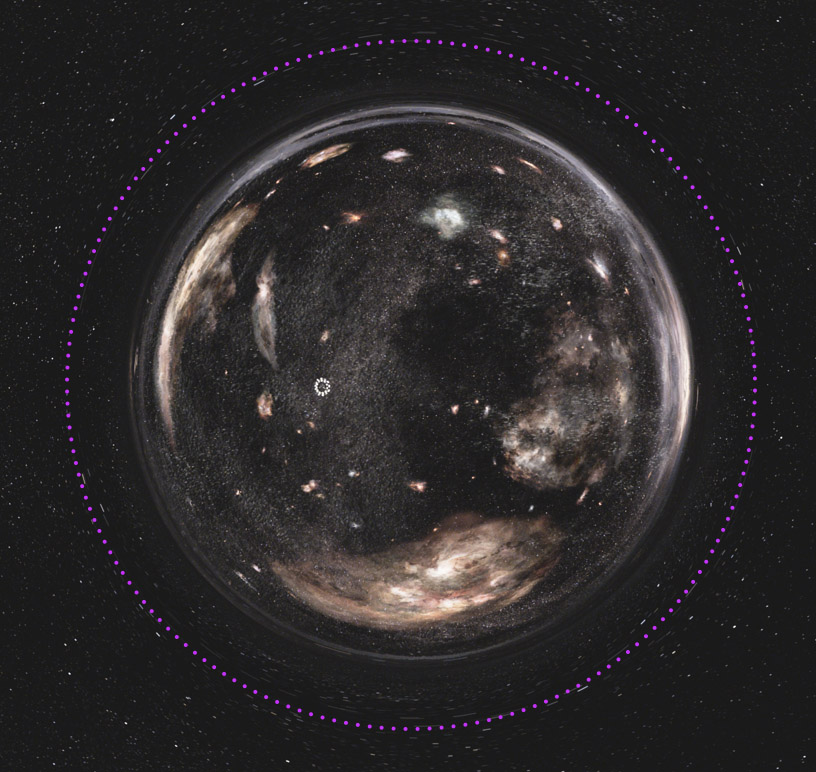}
\end{center}
\caption{An image of the distant galaxy seen through \emph{Interstellar}'s wormhole.  
The dotted pink circle is the wormhole's  
Einstein ring.  [From a trailer
for \emph{Interstellar}.  Created by
our Double Negative team. TM \& $\copyright$ Warner Bros. Entertainment Inc. (s15).
This image
may be used under the terms
of the Creative Commons Attribution-NonCommercial-NoDerivs 3.0 (CC BY-NC-ND
3.0) license. Any further distribution of these images must maintain attribution to the
author(s) and the title of the work, journal citation and DOI. You may not use the
images for commercial purposes and if you remix, transform or build upon the images,
you may not distribute the modified images.]  }
\label{fig:InterstellarWormhole}
\end{figure}

When we place this distant-galaxy image on the upper celestial sphere of Fig.\ \ref{fig:WH-TwoSides}
and place a simple star field on the lower celestial sphere, within which the camera
resides, then the moving camera sees the wormhole images shown in \emph{Interstellar}
and its trailers; for example, Fig.\  \ref{fig:InterstellarWormhole}.

\begin{figure}[h!]
\begin{center}
\includegraphics[width=1.0\columnwidth]{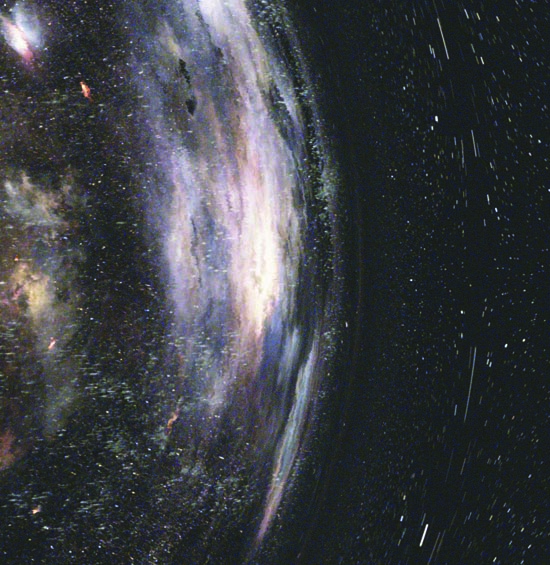}
\end{center}
\vskip-1.5pc
\caption{A close-up of \emph{Interstellar's} wormhole.
The long, streaked stars alongside the \emph{Einstein ring} are a result of motion blur:
the virtual camera's shutter is open for a fraction of a second (in this case, approximately 0.02 seconds)
during which the stars' lensed images appear to orbit the wormhole, causing the curved paths seen here.  [From \emph{Interstellar}, but cropped. Created by
our Double Negative team. 
TM \& $\copyright$ Warner Bros. Entertainment Inc. (s15).  This image
may be used under the terms
of the Creative Commons Attribution-NonCommercial-NoDerivs 3.0 (CC BY-NC-ND
3.0) license. Any further distribution of these images must maintain attribution to the
author(s) and the title of the work, journal citation and DOI. You may not use the
images for commercial purposes and if you remix, transform or build upon the images,
you may not distribute the modified images.] }
\label{fig:InterstellarWormholeCU}
\end{figure}

Students can create similar images, using their implementation of the map
described in the Appendix, and putting Fig.\  \ref{fig:InterstellarGalaxy} on the upper celestial
sphere.  They could be invited to explore how their images change as the camera
moves farther from the wormhole, closer, and through it, and as the wormhole
parameters are changed.

\subsection{The Einstein Ring}
\label{subsec:EinsteinRing}

Students could be encouraged to examine closely the changing image of the
wormhole in \emph{Interstellar} or one of its trailers, on a computer screen
where the student can move the image back and forth in slow motion.  Just
outside the wormhole's edge, at the location marked by a dotted circle
in Fig.\  \ref{fig:InterstellarWormhole}, the star motions (induced by
camera movement) are quite
peculiar.  On one side of the dotted circle, stars move rightward;
on the other, leftward.  The closer a star is to the circle, the faster it
moves; see Fig.\  \ref{fig:InterstellarWormholeCU}.  

The circle is called the wormhole's \emph{Einstein ring}.  This ring is actually the ring
image, on the camera's local sky, of a tiny light source that is precisely behind the wormhole
and on the same end of the wormhole as the camera.  
That location, on the celestial sphere and  precisely  opposite the camera, is actually
a \emph{caustic} (a singular, focal point) of the camera's past light cone.  As the camera orbits the
wormhole, causing this caustic to sweep very close to a star, the camera sees two images of the star, one just inside the Einstein ring and the other just outside it,  move rapidly around the ring in opposite directions.  This is the same behavior as occurs with the  Einstein ring of a black hole (see e.g.\ Fig.\ 
2 of our paper on black-hole lensing\cite{HoleLens}) and any other spherical gravitational lens, and it is also responsible for long, lenticular images of distant galaxies gravitationally lensed by a more nearby galaxy. \cite{bartelmann}

Students, having explored the wormhole's Einstein ring in a DVD or trailer of the movie, could be encouraged to go learn about Einstein rings and/or 
figure out for themselves how these 
peculiar star motions are produced.  They could then use their own implementation of our 
map to explore whether their explanation is correct. 

\section{Trip Through the Wormhole}
\label{sec:ThroughWormhole}

Students who have implemented the map (described in the Appendix) from 
the camera's local sky to the celestial spheres could be encouraged to explore, with
their implementation, what it looks like to travel through
the Dneg wormhole for various parameter values.  

We ourselves did so, together with Christopher
Nolan, as a foundation for \emph{Interstellar}'s wormhole trip.
Because the wormhole Nolan chose to visualize from the outside (upper left of
Fig.\ \ref{fig:DnegVary_a}; images in Figs.\ \ref{fig:InterstellarGalaxy} 
and \ref{fig:InterstellarWormholeCU}) is so short and
its lensing width so modest, the trip
was  quick and not terribly interesting, visually---not at all 
what Nolan wanted for his
movie.  So we generated additional through-the-wormhole clips for him, with
the wormhole parameters changed.  For a long wormhole, the trip was like traveling
through a long tunnel, too much like things seen in 
previous movies.  None of the clips, for any choice of parameters, had the compelling
freshness that Nolan sought.  

\begin{figure}[b!]
\begin{center}
\includegraphics[width=1.0\columnwidth]{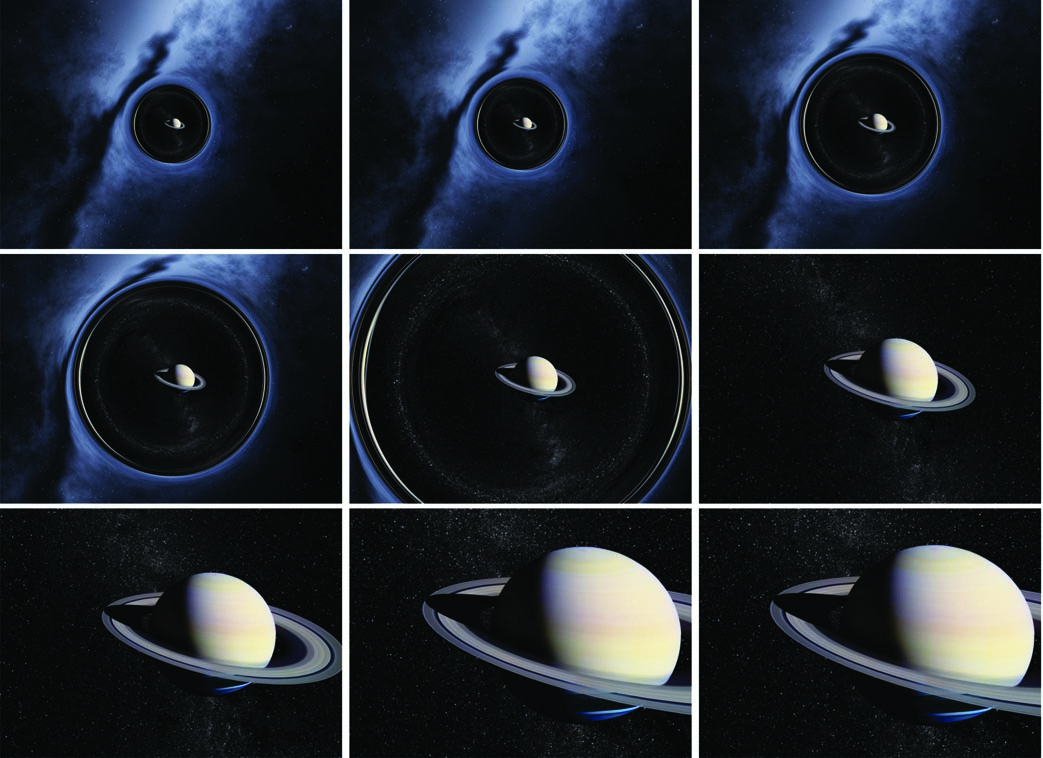}
\end{center}
\vskip-1.5pc
\caption{Still frames of a voyage through a short wormhole ($a/\rho=0.5)$ with
weak lensing ($\mathcal W/\rho = 0.05$), as computed with our DNGR code.}
\label{fig:WormholeTraversal}
\end{figure}

Moreover, none had the right  \emph{feel}. 
Figure \ref{fig:WormholeTraversal} illustrates this problem.
It shows stills from a trip through a moderately short
wormhole with $a/\rho = 0.5$ --- stills that students could replicate with 
their implementation.  
Although these images are interesting, the resulting animated sequence
is hard for an audience to interpret. The view of the wormhole appears to
scale up from its center, growing in size 
until it fills the
frame, and 
until none of the starting galaxy is visible; at this
point only the new galaxy can be seen, because we now are actually inside that new galaxy.
This is hard to interpret visually.  Because there is no parallax
or other relative motion in the frame, to the audience it looks 
like  the camera is zooming into the center of the wormhole using the camera's
zoom lens. In the visual grammar of filmmaking, this tells
the audience that we are zooming in for a closer look but  we are
still a distance from the wormhole; in reality we are travelling through
it, but this is not how it feels.

It was important for the audience to understand that the wormhole allows the Endurance to take a shortcut through the higher dimensional 
bulk.  To foster that understanding, Nolan asked the visual effects team to convey a sense of travel through an exotic environment, one that was thematically linked to the exterior appearance of the wormhole but 
also incorporated elements of passing landscapes and the sense of a rapidly approaching destination.  The visual effects artists at Double Negative combined existing DNGR visualisations of the wormhole's interior with layers of interpretive effects animation derived
from aerial photography of dramatic landscapes, adding lens-based photographic effects to tie everything in with the rest of the sequence.  The end result was a sequence of shots that told a story comprehensible by a general audience while resembling the wormhole's
interior, as simulated with DNGR.

\section{Conclusion}
\label{sec:Conclusion}

As we wrote this paper, we became more and more enthusiastic about the
educational opportunities provided by our \emph{Interstellar} experience.  The
tools we used in building, scoping out, and exploring \emph{Interstellar}'s 
wormhole---at least those discussed in this paper---should be easily accessible
to fourth year undergraduates studying relativity, as well as to graduate students.
And the movie itself, and our own route to the final wormhole images in the
movie, may be a strong motivator for students.

\appendix*

\section{The Ray-Induced Map from the Camera's Local Sky to the Two Celestial Spheres}

In this appendix we describe our fairly simple procedure for generating the map
from points $\{\theta_{\rm cs},\phi_{\rm cs}\}$ on the camera's local sky to points
$\{\theta',\phi',s\}$ on the wormhole's celestial sphere, with $s=+$ for the upper
celestial sphere and $s=-$ for the lower.

\subsection{The Ray Equations}

As we discussed in Sec.\ \ref{subsec:MapFoundations}, the map is generated by light rays that travel backward
in time from the camera to the celestial spheres.  In the language of general relativity,
these light rays are null (light-like) geodesics and so are 
solutions of the geodesic equation 
\begin{equation}
{d^2 x^\alpha\over d\zeta^2} + {\Gamma^\alpha}_{\mu\nu}
{dx^\mu\over d\zeta}{dx^\nu\over d\zeta}=0\;.
\label{eq:GeodesicEqn}
\end{equation}
Here the ${\Gamma^\alpha}_{\mu\nu}$
are Christoffel symbols (also called connection coefficients) that are constructable
from first derivatives of the metric coefficients, and $\zeta$ is the so-called \emph{affine
parameter}, which varies along the geodesic.

This form of the geodesic equation is fine for analytical work, but for numerical work it is best rewritten in the language of Hamiltonian mechanics.
Elsewhere \cite{PriceThorne} one of us will discuss, pedagogically, the advantages and 
the underpinnings of this Hamiltonian rewrite.  

There are several different Hamiltonian formulations of the geodesic equation. The one we advocate is  sometimes called the ``super-Hamiltonian'' because of its beauty and power, but we will stick to the usual word ``Hamiltonian''.  The general formula for this Hamiltonian is\cite{PriceThorne,MTW} 

\begin{equation}
H(x^\alpha, p_\beta) = \frac12 g^{\mu\nu}(x^\alpha) p_\mu p_\nu\; .
\label{GeneralHamiltonian}
\end{equation}
Here $g^{\mu\nu}$ are the contravariant components of the metric, $x^\alpha$ is the
coordinate of a photon traveling along the ray,  and $p_\alpha$ is the generalized momentum  that is canonically conjugate to  $x^\alpha$ and it turns out to be the same as the covariant component
of the photon's 4-momentum. 
Hamilton's equations, with the affine parameter $\zeta$ playing the role of time, take the standard form
\begin{subequations}
\begin{eqnarray}
{dx^\alpha \over d\zeta} &=& {\partial H\over\partial p_\alpha} = g^{\alpha \nu} p_\nu\;,
\label{Hamiltonianx} \\
{dp_\alpha \over d\zeta} &=& - {\partial H\over\partial x^\alpha} = 
- \frac12 {\partial{g^{\mu\nu}}\over \partial x^\alpha} p_\mu p_\nu\;.
\label{Hamiltonianp}
\end{eqnarray}
\label{HamiltonEqns}
\end{subequations}

In the first of Eqs.\ (\ref{HamiltonEqns}), the metric raises the index on the covariant momentum,
so it becomes $p^\alpha = dx^\alpha/d\zeta$, an expression that may be familiar
to students.  The second expression may not be so familiar, but it can be given
as an exercise for students to show that the second equation, together with 
$p^\alpha = dx^\alpha/d\zeta$, is equivalent to the usual form
(\ref{eq:GeodesicEqn}) of the geodesic equation.

For the general wormhole metric (\ref{eq:GeneralMetric}), the superhamiltonian 
(\ref{GeneralHamiltonian}) has the simple form
\begin{equation}
H = \frac12\left[-p_t^2+p_\ell^2+{p_\theta^2\over r(\ell)^2} + 
{p_\phi^2\over r(\ell)^2 \sin^2\theta}\right]\;.
\label{eq:Hamiltonian}
\end{equation}

Because this superhamiltonian is independent of the time coordinate $t$ and of the azimuthal
coordinate $\phi$, $p_t$ and $p_\phi$ are conserved along a ray
[cf.\ Eq.\ (\ref{Hamiltonianp})].  Since $p^t = dt/d\zeta = -p_t$, changing the numerical value of $p_t$ merely renormalizes the affine parameter $\zeta$; so without  loss of
generality, we set $p_t=-1$, which implies that $\zeta$ is 
equal to time $t$ [Eq.\ (\ref{eq:REt}) below]. Since photons travel at the speed of light, $\zeta$ is also distance 
travelled (in our geometrized units where the speed of light is one).  

We use the notation $b$ for the conserved quantity $p_\phi$: 
\begin{subequations}
\begin{equation}
b=p_\phi\;.
\label{eq:bdef}
\end{equation}  
Students should easily be able to show that, because we set $p_t=-1$, 
this $b$ is the ray's impact parameter relative to the (arbitrarily chosen\cite{PolarAxis}) 
polar axis.  Because the wormhole is spherical, there is a third conserved
quantity for the rays, its total angular momentum, which (with $p_t=-1$) is the same as its
impact parameter $B$ relative to the hole's center
\begin{equation}
B^2= p_\theta^2+{p_\phi^2 \over\sin^2\theta}\;.
\label{eq:B2}
\end{equation}
\end{subequations}

By evaluating Hamilton's equations for the wormhole Hamiltonian 
(\ref{eq:Hamiltonian}) and inserting the conserved quantities on the right-hand side,
we obtain the following ray equations:
\begin{equation}
{dt\over d\zeta} = -p_t = 1\;, \label{eq:REt} 
\end{equation}
which reaffirms that $\zeta = t$ (up to an additive constant); and, replacing $\zeta$ by $t$:
\begin{subequations}
\begin{eqnarray}
{d\ell\over dt} &=& p_\ell\;, \label{eq:RExi} \\
{d\theta\over dt} &=& {p_\theta\over r^2} \;, \label{eq:REtheta} \\
{d\phi\over dt} &=& {b\over r^2 \sin^2\theta}\\
{d p_\ell\over dt}&=& B^2{d r /d\ell \over r^3}\;, \label{eq:REpxi} \\
{d p_\theta\over dt} &=& {b^2\over r^2} {\cos\theta \over \sin^3 \theta}\;.
\label{eq:REptheta}
\end{eqnarray}
\label{eq:RE}
\end{subequations}

These are five equations for the five quantities $\{\ell,\theta,\phi,p_\ell,
p_\theta\}$ as functions of $t$ along the geodesic (ray).

It is not at all obvious from these equations, but they guarantee (in view of spherical
symmetry) that the lateral (nonradial) part of each ray's motion is along a great circle.

These equations may seem like an overly complicated way to describe a ray.  
Complicated, maybe; but near ideal for simple numerical integrations.  They are stable
and in all respects well behaved everywhere except the poles $\theta=0$ and
$\theta = \pi$, and they are easily implemented in student-friendly software such
as Mathematica, Maple and Matlab.

\subsection{Procedure for Generating the Map}

It is an instructive exercise for students to verify the following procedure for constructing
the map from the camera's local sky to the two celestial spheres:

\begin{enumerate}
\item
Choose a camera location $(\ell_c, \theta_c, \phi_c)$.  It might best be on the equatorial plane, $\theta_c=\pi/2$, so the coordinate singularities at $\theta=0$ and $\theta = \pi$ are as far from the camera as possible.  

\item
Set up local Cartesian coordinates centered on the camera, with $x$ along the
direction of increasing $\ell$ (toward the wormhole on the Saturn side; away
from the wormhole on the Gargantua side), $y$ along the direction  of increasing $\phi$, and
$z$ along the direction of \emph{decreasing} $\theta$, 
\begin{equation}
\mathbf e_x = \mathbf e_{\hat \ell} \;, \quad
\mathbf e_y = \mathbf e_{\hat \phi}\;, \quad 
\quad \mathbf e_z= - \mathbf e_{\hat \theta}\;.
\label{eq:bases}
\end{equation}  
Here $\mathbf e_{\hat \ell}$, $\mathbf e_{\hat \theta}$ and $\mathbf e_{\hat \phi}$ are
unit vectors that point in the $\ell$, $\theta$, and $\phi$ directions.  (The hats tell us
their lengths are one.)
Figure \ref{fig:WH-TwoSides} shows these camera basis vectors, for the special 
case where the camera is in the equatorial plane.  The minus sign in our choice 
$\mathbf e_z= - \mathbf e_{\hat \theta}$ makes the camera's $e_z$ parallel to the 
wormhole's polar axis on the Gargantua side of the wormhole, where $\ell$ is 
positive. 

\item
Set up a local spherical polar coordinate system for the camera's local sky in the usual
way, based on the camera's local Cartesian coordinates; cf.\ 
Eq.\ (\ref{eq:N}) below.

\item
Choose a direction $(\theta_{cs}$, $\phi_{cs})$ on the camera's local sky.  The unit
vector $\mathbf N$ pointing in that direction has Cartesian components
\begin{subequations}
\begin{eqnarray}
N_x &=&\sin\theta_{cs} \cos\phi_{cs}\;, \quad
N_y= \sin\theta_{cs} \sin\phi_{cs}\;, \nonumber \\
N_z &=& \cos\theta_{cs}\;.
\label{eq:N}
\end{eqnarray}
Because of the relationship (\ref{eq:bases}) between bases, 
the direction $\mathbf n$ of propagation of the incoming ray that arrives from  
direction $-\mathbf N$,
has components in the global spherical polar basis
\begin{equation}
n_{\hat\ell} = -  N_x\;,\quad
n_{\hat\phi} = - N_y\;,  \quad
n_{\hat\theta} = + N_z\;. 
\label{eq:n}
\end{equation}
\item
Compute the incoming light ray's canonical momenta from
\begin{equation}
p_\ell = n_{\hat\ell}\;, \quad
p_\theta = r n_{\hat\theta}\;, \quad
p_\phi = r \sin\theta n_{\hat\phi}\;
\label{eq:ps}
\end{equation}
(it's a nice exercise for students to deduce these equations from
the relationship between the covariant components of the
photon 4-momentum and the components on the unit basis
vectors). 
Then compute the ray's constants of motion from
\begin{eqnarray}
b&=&p_\phi = r \sin\theta n_{\hat\phi}\;, \nonumber \\
B^2&=&p_\theta^2 + {p_\phi^2 \over \sin^2\theta} = r^2 ( n_{\hat\theta}^2 +  n_{\hat\phi}^2)\;.
\label{eq:constants}
\end{eqnarray}
\end{subequations}
\item
Take as initial conditions for ray integration that at $t=0$ 
the ray begins at the camera's location, $(\ell,\theta,\phi)=(\ell_c,\theta_c,\phi_c)$
with canonical momenta (\ref{eq:ps}) and constants of motion (\ref{eq:constants}).
Numerically integrate the ray equations (\ref{eq:RE}), subject to these initial conditions,
from $t=0$ backward along the ray to time $t_i=-\infty$ (or some extremely negative, finite
initial time $t_i$).
If $\ell(t_i)$ is negative, then
the ray comes from location $\{\theta',\phi'\} = \{\theta(t_i), \phi(t_i)\}$
on the Saturn side of the wormhole, $s=-$.  If $\ell(t_i)$ is positive, then
the ray comes from location $\{\theta',\phi'\} = \{\theta(t_i), \phi(t_i)\}$
on the Gargantua side of the wormhole, $s=+$.
\end{enumerate}

\subsection{Implementing the map}
\label{subsec:MapImplementation}

Evaluating this map numerically should be a moderately easy task for students.

Kip Thorne, the author among us who is a total klutz at numerical work, did it using
Mathematica, and then used that map---a numerical table of $\{\theta',\phi',s\}$
as a function of $\{\theta_{\rm cs}, \phi_{\rm cs}\}$---to make camera-sky images
of whatever was placed on the two celestial spheres.  For image processing,
Thorne first built an interpolation of the map using the Mathematica command
\textsc{ListInterpolation}; and he then used this interpolated map, together with
Mathematica's command \textsc{ImageTransformation}, to produce the
camera-sky image from the images on the two celestial spheres.

\begin{acknowledgments}

For extensive advice on our wormhole visualizations, we thank
Christopher Nolan.   
For contributions to DNGR and its wormhole applications, we thank  members of the Double Negative R\&D team
Sylvan Dieckmann, Simon Pabst, Shane Christopher, Paul-George Roberts, and Damien Maupu; and also Double Negative artists 
Zoe Lord, Fabio Zangla, Iacopo di Luigi, Finella Fan, Tristan Myles, Stephen Tew, and Peter Howlett.

The construction of our code DNGR was funded by Warner Bros. Entertainment Inc., for generating visual effects for the movie \emph{Interstellar}.  We thank Warner Bros. for authorizing this code's additional use for scientific research and physics education, and in particular the
work reported in this paper.

\end{acknowledgments}


\begin{thebibliography}{99}


\bibitem{Contact}
Carl Sagan, \textit{Contact} (Simon and Schuster, New York, 1985).

\bibitem{ContactMovie}
\textit{Contact}, The Movie, directed by Robert Zemeckis ($\copyright$ Warner Bros., 1997).

\bibitem{MorrisThorne} Michael S. Morris and Kip S. Thorne, ``Wormholes in 
spacetime and their use for interstellar travel: A tool for teaching general relativity,''
\textit{Am.\ J.\ Phys.} \textbf{56} 395--412 (1988).

\bibitem{Interstellar}
\textit{Interstellar}, directed by Christopher Nolan, screenplay by Jonathan
Nolan and Christopher Nolan ($\copyright$ Warner Bros, 2014).


\bibitem{TSI} Kip Thorne, \textit{The Science of Interstellar} (W.W.\  Norton
and Company, New York, 2014).

\bibitem{EverettRoman} Allen Everett and Thomas Roman, \textit{Time Travel and
Warp Drives} (University of Chicago Press, Chicago, 2012).

\bibitem{FriedmanHiguchi} John L. Friedman and Atsushi Higuchi ``Topological 
censorship and chronology protection,''  \textit{Annalen Phys.} {\bf 15}, 109--128 (2006).

\bibitem{Lobo} Francisco S.\ N.\ Lobo ``Exotic solutions in general relativity: traversable wormholes and `warp drive' spacetimes,'' {\it Classical and Quantum Gravity Research 5 Progress} (Nova Science Publishers, Hauppauge, NY, 2008), 1--78.

\bibitem{MTU} Michael S. Morris, Kip S. Thorne, and Ulvi Yurtsever, ``Wormholes, time machines, and the weak energy condition," \textit{Phys.\ Rev.\ Lett.}, {\bf 61}, 1446-1449 (1988).

\bibitem{OtherStars} See, e.g., chapter 13 of \textit{The Science of Interstellar}\cite{Interstellar}.

\bibitem{Hartle} James B.\ Hartle, \textit{Gravity: An Introduction to Einstein's
General Relativity} (Addison Wesley, San Francisco, 2003).

\bibitem{HoleLens}
Oliver James, Eug\'enie von Tunzelmann, Paul Franklin, and Kip S.\ Thorne, ``Gravitational lensing by spinning black holes in astrophysics, and in the movie \textit{Interstellar}'', \emph{Class.\  Quant.\ Grav.}, {\bf 32}, 065001 (2015).

\bibitem{Ellis} Homer G. Ellis, ``Ether flow through a drainhole: a particle model in general relativity'', \textit{J.\ Math.\ Phys.} {\bf 14}, 104--118 (1973).

\bibitem{Ellis-MT} Fifteen years later, Morris and Thorne\cite{MorrisThorne} wrote
down this same metric, among others, and being unaware of Ellis's paper, failed to
attribute it to him, for which they apologize.  Regretably, it is sometimes called the Morris-Thorne wormhole metric.

\bibitem{Muller} Thomas M\"uller, ``Visual appearance of a Morris-Thorne-Wormhole,''
ÊÊ \textit{Am. J. Phys.} {\bf 72}, 1045--1050 (2004), which is based in part on
Daniel Weiskopf, ``Visualization of four-dimensional spacetimes,'' PhD thesis at 
der Eberhard-Karls-Universit\"at zu T\"ubingen, available at
\url{http://nbn-resolving.de/urn:nbn:de:bsz:21-opus-2400}.  
Wormhole images based
on M\"uller's paper are available at 
 \url{https://www.youtube.com/watch?v=FHmupoY4nZU&index=2&list=PLdcIglDT8_FEnv0MmGrb04azONGpgqbq7}, and 
 also in the paper Hans Ruder et.\ al., ``How computers can help us in creating an intuitive access to relativity,'' \textit{New J.\ Phys.} {\bf 10}, 125014 (2008).  For a movie by Covin Zahn of what
 it looks like to travel through an Ellis wormhole, see 
\url{http://www.spacetimetravel.org/wurmlochflug/wurmlochflug.html}.


\bibitem{GLRL} Tommaso Treu, Philip J.\ Marshall and Douglas Clowe, ``Resource
Letter GL-1: Gravitational Lensing,''  \textit{Am.\ J.\ Phys.} {\bf 80}, 753--763 (2012);
http://arxiv.org/pdf/1206.0791v1.pdf and  https://groups.diigo.com/group/gravitational-lensing.

\bibitem{HartleEmbed} This is the same as Eq.\ (7.46b) 
of Hartle,\cite{Hartle} where, however, our $\ell$ is denoted $\rho$.

\bibitem{NolanRadius}  See the technical notes for chapter 15 of
\emph{The Science of Interstellar},\cite{TSI} pages 294--295.

\bibitem{LensingLength}  From the embedding equation (\ref{eq:EmbeddingEqn}) and
$dr/d\ell = (2/\pi) \arctan(2\ell/\pi\mathcal M)$ [Eq.\ (\ref{eq:rellDneg})],  it follows that $\mathcal W/\mathcal M =
- \ln[\sec(\pi/2\sqrt2)] + (\pi/2\sqrt2)\tan(\pi/2\sqrt2) = 1.42053... $.

\bibitem{Malmer} Mattias Malmer, http://apod.nasa.gov/apod/ap041225.html .

\bibitem{BLINN} J.\ F.\  Blinn  and M.\ E.\ Newell, ``Texture and reflection in computer generated images,'' \textit{Communications of the ACM} \textbf{19},  
542--547 (1976).

\bibitem{bartelmann} M.\ Bartelmann  ``Gravitational lensing,'' {\it Class.\ Quant.\ Grav.} {\bf 27} 233001 (2010)


\bibitem{PriceThorne} Richard H.\ Price and Kip S.\ Thorne, ``Superhamiltonian
for geodesic motion and its power in numerical computations,'' \textit{Amer.\ J.\ Phys.},
in preparation.

\bibitem{MTW} Section 21.1 of Charles W.\ Misner, Kip S.\ Thorne and John Archibald Wheeler,
\emph{Gravitation} (W.\ H.\ Freeman, San Francisco, 1973).

\bibitem{PolarAxis} The polar axis is arbitrary because the wormhole's geometry is
spherically symmetric.


\end{thebibliography}
\end{document}